\documentclass[aps,prl,superscriptaddress,floatfix,amsmath,amsfonts,twocolumn]{revtex4-2}  
\usepackage[english]{babel}

\usepackage{enumitem}
\usepackage{amsmath}
\usepackage{graphicx}
\usepackage{epsfig}
\usepackage{bm}
\usepackage{bbold}
\usepackage{amssymb}
\usepackage{color}
\usepackage[normalem]{ulem}  %

\begin{document}

\title{Spin resonance induced by a mechanical rotation of a polariton condensate} 

\author{A.V. Yulin}
\affiliation{Department of Physics, ITMO University, Saint Petersburg 197101, Russia}
\affiliation{Science Institute, University of Iceland, Dunhagi 3, IS-107, Reykjavik, Iceland}

\author{I.A. Shelykh}
\affiliation{Science Institute, University of Iceland, Dunhagi 3, IS-107, Reykjavik, Iceland}
\affiliation{Department of Physics, ITMO University, Saint Petersburg 197101, Russia}

\author{E. S. Sedov}
\affiliation{Russian Quantum Center, Skolkovo, Moscow 143025, Russia}
\affiliation{Spin-Optics laboratory, St. Petersburg State University, St. Petersburg 198504, Russia}
\affiliation{Vladimir State University, Vladimir 600000, Russia}

\author{A.V. Kavokin}
\affiliation{Westlake University, School of Science, 18 Shilongshan Road, Hangzhou 310024, Zhejiang Province, China}
\affiliation{Westlake Institute for Advanced Study, Institute of Natural Sciences, 18 Shilongshan Road, Hangzhou 310024, Zhejiang Province, China}
\affiliation{Spin-Optics laboratory, St. Petersburg State University, St. Petersburg 198504, Russia}

\date{\today}

\begin{abstract}
We study theoretically the polarization dynamics in a ring-shape bosonic condensate of exciton-polaritons confined in a rotating trap. The interplay between the rotating potential and TE-TM splitting of polariton modes offers a tool of control over the spin state and the angular momentum of the condensate. Specific selection rules describing the coupling of pseudospin and angular momentum are formulated. The resonant coupling between states having linear and circular polarizations leads to the polarization beats. The effect may be seen as a polariton analogy to the electronic magnetic resonance in the presence of constant and rotating magnetic fields. Remarkably, spin beats are induced by a purely mechanical rotation of the condensate.  

\end{abstract}

\maketitle

\textit{Introduction}. Exciton-polaritons are hybrid light-matter quasiparticles emerging in the regime of the strong coupling between a photonic mode of a planar semiconductor microcavity and an excitonic resonance in a quantum well embedded in the antinode of a cavity mode.
From their photonic component polaritons inherit extremely small effective mass (about $10^{-5}$ of the mass of free electrons) and large coherence length (in the mm scale) \cite{Ballarini2017}.
On the other hand, the presence of an excitonic component leads to the sensitivity of the polariton systems to external electric and magnetic fields, and robust polartion-polariton interactions \cite{Glazov2009}.

Remarkable tunability of cavity polaritons allows to engineer their spatial confinement in a variety of experimental geometries, ranging from individual micropillars \cite{Bajoni2008,Ctistis2010,Ferrier2011,Real2021} to the systems of several coupled pillars forming so-called polariton molecules \cite{Galbiati2012,Sala2015} or periodically arranged arrays of the pillars forming polariton superlattices \cite{Milicevic2017,Suchomel2018,Whittaker2018,Whittaker2021,Kuriakose2022}. Annular geometries are of particular interest, as in this case the interplay between non-trivial topology of the system and polarization TE-TM and Zeeman splittings can lead to a variety of intriguing physical phenomena, such as formation of the polaritonic persistent currents \cite{Lukoshkin2018} including symmetry breaking in spinor polariton current states~\cite{SciRep1122382}, linear \cite{Shelykh2010} and nonlinear \cite{Zezyulin2018} polaritonic Aharonov-Bohm effect, topological spin Meissner effect \cite{Gulevich2016}, angular momentum fractionalization \cite{Sedov2022}, and others. Moreover, it was recently proposed, that polariton rings can form a material platform for the realization of optical qubits \cite{Xue2021}.

\begin{figure}[tb!]
\begin{center}
\includegraphics[width=\linewidth]{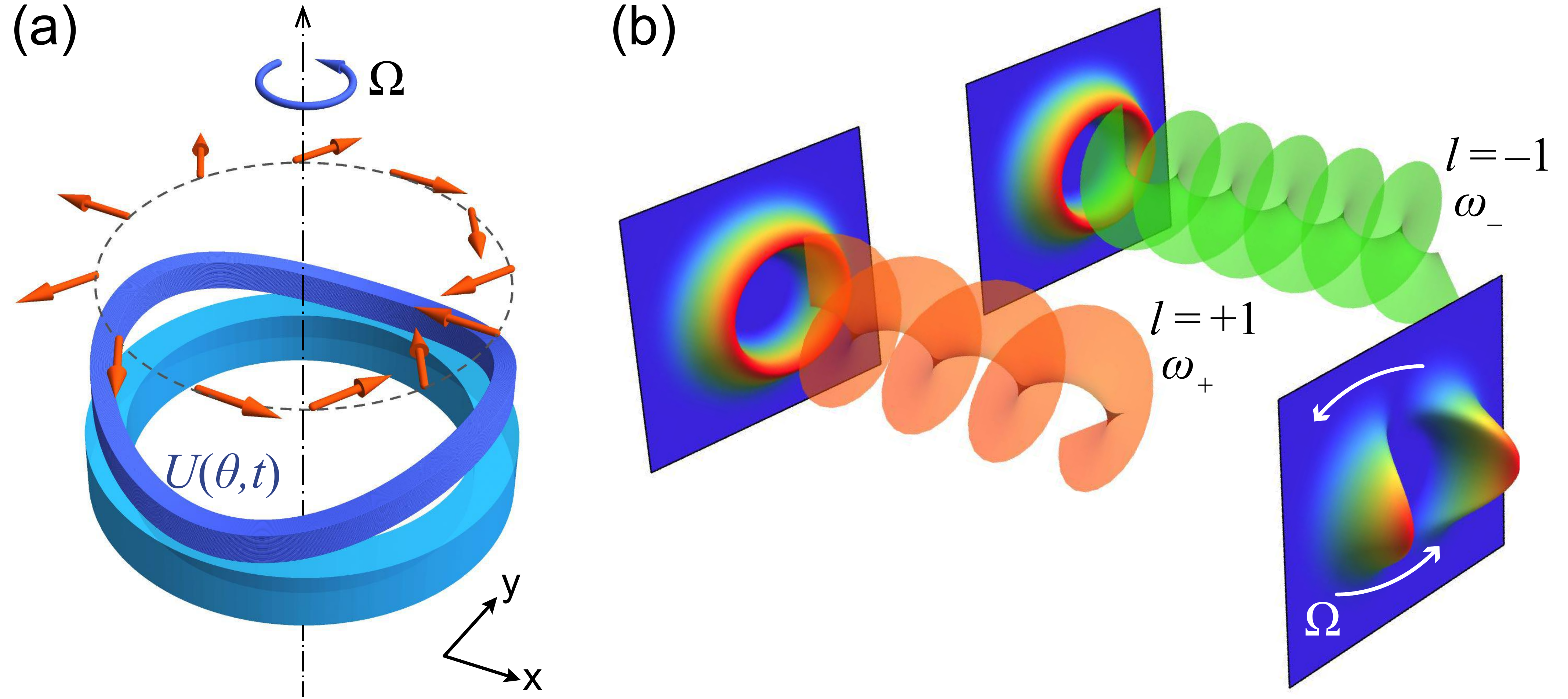}
\end{center}
\caption{ \label{FIG_Scheme}
 (a) Polariton ring condensate (blue) in the presence of TE-TM splitting, producing an effective magnetic field (orange arrows) acting on pseudospins of polaritons, subjected to an external rotating perturbation potential ${U(\theta,t)=U_0\cos(2\theta-\Omega t)}$. Orange arrows indicate the directions of the effective magnetic field produced by the TE-TM splitting along the ring. (b). Superposition of the two Laguerre-Gaussian laser beams characterised by the angular momenta $l=\pm1$ and slightly different in frequencies $\omega_{\pm}$.  The frequency detuning between the two beams leads to the appearance of the angular potential, rotating with the frequency $\Omega=\omega_+-\omega_-$.}
\end{figure}

In the present paper we theoretically predict a strong polarization resonance to appear in a ring-shape polariton condensate subject to a rotating potential trap. Such rotating traps can be produced by optical pumping with Laguerre-Gaussian laser beams as detailed below. We demonstrate, that linear to circular polarization coupling provided by the perturbation leads to the strong beats between the corresponding states. The phenomenon is a polaritonic counterpart of the magnetic resonance experienced by the spin of an electron placed in a combination of constant and rotating magnetic fields.

\textit{The model.} We assume the geometry of an experiment illustrated in Fig.~\ref{FIG_Scheme}(a). A thin polariton ring of the radius $R$, is subjected to the external scalar perturbation potential having the form 
\begin{equation}
    U(\theta,t)=U_0\cos(2\theta-\Omega t),
\end{equation}
where $\theta$ is an angular coordinate along the ring.We assume that the thickness of the ring $d\ll R$, so that only the lowest radial mode can be excited. Such kind of a perturbation results from a superposition of two optical Laguerre-Gaussian modes having the angular momenta $l=\pm1$, which are slightly detuned in energy~\cite{lagoudakis2022} (see Fig.~\ref{FIG_Scheme}(b)).

\begin{figure}[tb!]
\begin{center}
\includegraphics[width=\linewidth]{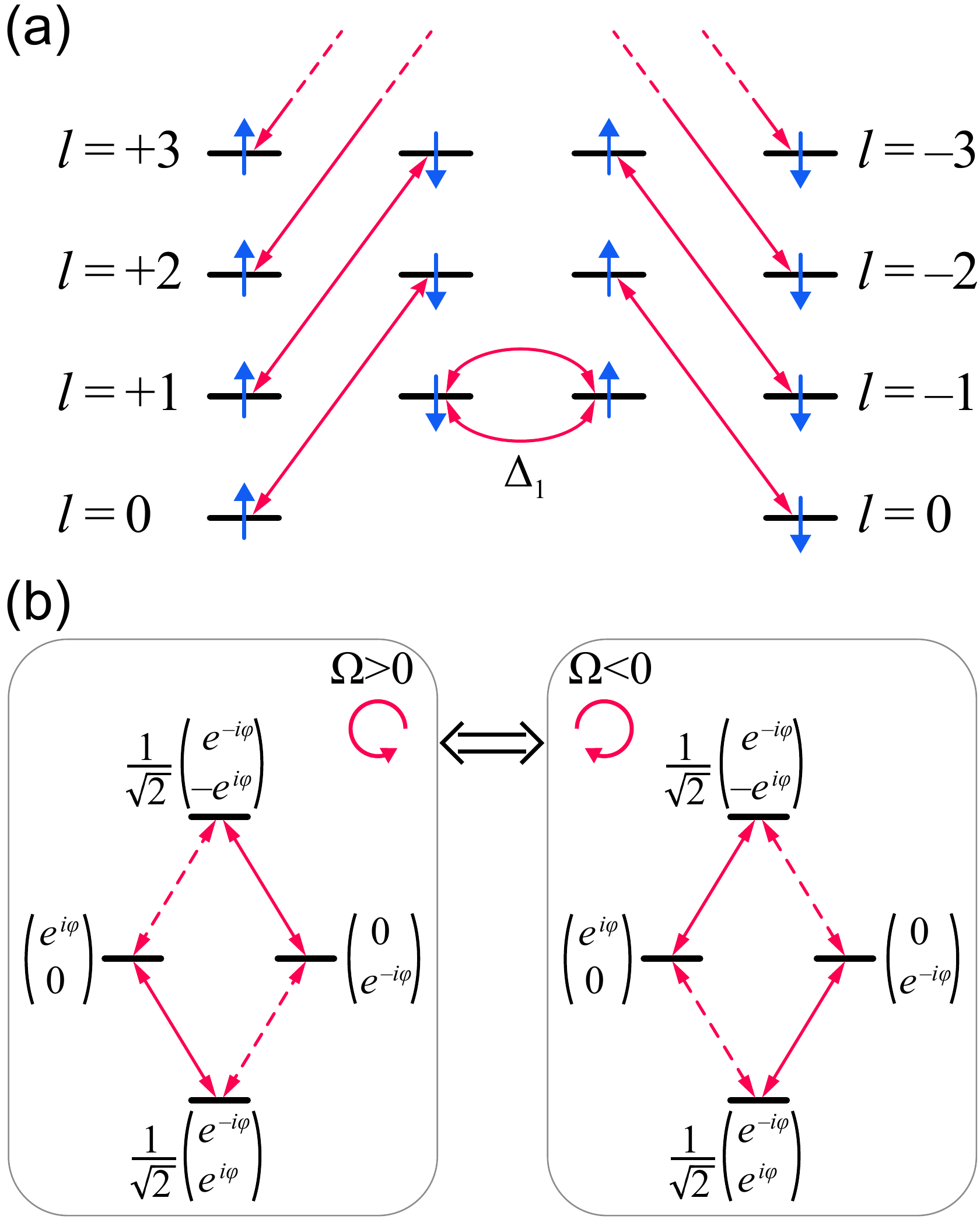}
\end{center}
\caption{ \label{FIG_Diags}
(Color online)
(a) Scheme of the coupling of the levels of a polariton ring by TE-TM interaction $\Delta_1$ (see Eq.~\eqref{H0}). Black horizontal lines correspond to the energy levels for the case $\Delta_1$=0, up and down vertical arrows denote states with right and left circular polarizations, $l$ correspond to the winding numbers.  The ground state is twice degenerate in polarization, all other states are four time degenerate: in polarization and sign of $l$. If $\Delta_1 \neq 0$, the states with $l_{\downarrow}-l_{\uparrow}=2$ become mixed, and degenaracies are partially lifted. Among all coupled pairs, the one corresponding to the states with $l_{\downarrow}=1,l_{\uparrow}=-1$ is particular, as corresponding states have equal energies, and thus they are most efficiently coupled by TE-TM splitting. (b). The scheme illustrating the transitions induced by the rotating potential $ U(\theta,t)=U_0\cos(2\theta-\Omega t) $ in the subspace of the states with winding numbers $l_{\uparrow,\downarrow}=\pm1$. Upper and lower states are strongly split due to TE-TM interaction and are linearly polarized in radial and tangential directions, respectively. Two states in the middle remain degenerate in energy and correspond to two circular polarizations. Resonant and anti-resonant transitions are shown by solid and dashed red arrows, respectively. }
\end{figure}

The state of the system is described by a two component spinor, corresponding to the two opposite circular polarizations $\psi=(\psi_\uparrow,\psi_\downarrow)^{\text{T}}$, and its dynamics is given by a Shrödinger-type equation,
\begin{equation}
    i\hbar\partial_t\psi=\hat{H}\psi.
    \label{Schrodinger}
\end{equation}

In this paper, we focus on the conservative linear limit, where we neglect the dissipative nature of cavity polaritons and polariton-polariton interactions. These approximations will allow us to reveal the proposed effect analytically. We acknowledge that the optical pump in polariton systems normally crates a complex potential whose imaginary part corresponds to the effective gain seen by the polaritons, but related effects require special consideration which is left for the follow-up paper. In the same time we would like to emphasise that the conservative case is also very relevant from physical point of view. 

The operator $\hat{H}$ in this approximation is hermitian and corresponds to the Hamiltonian of the system, which reads \cite{Kozin2018}:
\begin{equation}
    \hat{H}=\hat{H}_0+U(\theta,t),
\end{equation}
where, in the basis of the circular polarizations
\begin{eqnarray}
\hat{H}_0=\frac{\hbar^2}{2m_pR^2}\left(\begin{array}{cc}-\partial_\theta^2 & \Delta_1 e^{-2i\theta} \\ \Delta_1 e^{2i\theta} & -\partial_\theta^2
\end{array}\right).
\label{H0}
\end{eqnarray}
Here $m_p$ is an effective mass of polaritons, $\partial_\theta=d/d\theta$, and the dimensionless parameter $\Delta_1$ is proportional to the inverse square of the ring thickness $d$ and characterizes the value of the TE-TM splitting in the system.

To better understand the effect proposed here qualitatively, let us first consider the structure of the energy levels of a ring described by the Hamiltonian~\eqref{H0}. In the case where TE-TM splitting is absent ($\Delta_1=0$), the states of the opposite circular polarizations are decoupled from each other, so that the energy levels are characterized by the independent winding numbers $l_{\uparrow}$ and $l_{\downarrow}$, corresponding to right and left circular polarizations, respectively. The ground state with $l_{\uparrow}=l_{\downarrow}=0$ is twice degenerated, while all upper energy levels are degenerated four times, as clockwise and anticlockwise rotations, corresponding to the different signs of $l$ are all equivalent. 

The presence of the TE-TM splitting mixes the states with opposite circular polarizations having the winding numbers $l_- - l_+=2$, as it is shown by the red arrows in Fig.~\ref{FIG_Diags}(a). One can see that the states with $l_{\pm}=\pm 1$ are different from all the rest \cite{Rubo2022}, as only within this quadruplet we have a pair of the states with the same energy coupled to each other. We will therefore focus on them and project the dynamic equation~\eqref{Schrodinger} into the corresponding subspace. Its four basis vectors split in the two groups. 

The first one corresponds to the states with ${l_{\uparrow}=-1}$, ${l_{\downarrow}=+1}$, coupled by TE-TM splitting. As the result, two linearly polarized states, with tangential and radial polarizations are produced, with the energies ${E_T=\hbar^2(1-\Delta_1)/2m_pR^2}$ and ${E_R=\hbar^2(1+\Delta_1)/2m_pR^2}$ and wave functions
\begin{eqnarray}
\psi_T=\frac{1}{\sqrt{2}}\left(\begin{array}{c}e^{-i\theta} \\ e^{i\theta}\end{array}\right),\psi_R=\frac{1}{\sqrt{2}}\left(\begin{array}{c}e^{-i\theta} \\ -e^{i\theta}\end{array}\right).
\end{eqnarray}

The second group corresponds to a pair of degenerate states, which stem from the states with ${l_{\uparrow}=+1}$ and ${l_{\downarrow}=-1}$, with some admixture of the states with $l_\uparrow=-3,l_\downarrow=3$. For simplicity, we neglect this admixture, assuming that ${\Delta_1\ll 1}$. In this case, these states remain circularly polarized, their energies being ${E_c=\hbar^2/2m_p R^2}$ and the wave functions
\begin{eqnarray}
\psi_\uparrow=\left(\begin{array}{c}e^{i\theta} \\ 0\end{array}\right),\psi_\downarrow=\left(\begin{array}{c} 0 \\ e^{-i\theta}\end{array}\right).
\end{eqnarray}

Approximating the wavefunction as $\psi=\sum_jA_j\psi_j$, where $j=T,R,\uparrow,\downarrow$ we obtain the following set of the coupled equations for the amplitudes of the modes $A_j$:
\begin{subequations}
	\label{eq:coupled_modes_A}
\begin{eqnarray}	
\partial_t A_{T}=-i\omega_c(1-\Delta_1) A_T+i \eta \left( e^{i \Omega t} A_{\uparrow} +e^{-i \Omega t} A_{\downarrow}\right), \, \\
\partial_t A_{R}=-i\omega_c(1+\Delta_1) A_R+ i \eta \left( e^{i \Omega t} A_{\uparrow} -e^{-i \Omega t} A_{\downarrow}\right), \, \\
\partial_t A_{\uparrow}=-i\omega_c A_{\uparrow}+i \eta e^{-i \Omega t} (A_{T}+A_{R}), \,  \\
\partial_t A_{\downarrow}=-i\omega_c A_{\downarrow}+i \eta e^{i \Omega t} (A_{T}-A_{R}), \,
\end{eqnarray}
\end{subequations}
where $\omega_c=E_c/\hbar,\eta=U_0/\hbar$. As one can see, the  rotating perturbation mixes the states with linear and circular polarizations, while it does not mix the states of the opposite linear polarizations $A_R$ and $A_T$ directly. This is because of the interplay between TE-TM splitting and particular symmetry of the perturbation, which mixes the components with the winding numbers differing by two. 

\textit{Rotating wave approximation.}
Let us consider the resonant case, where $\Omega \approx \omega_c\Delta_1$. One can see that the couplings can be either resonant, or antiresonant, depending on the sign of  $\Omega$, i.e. the direction of the rotation of the perturbation. In the case, when $\Omega>0$ the tangentially linearly polarized lowest energy state resonantly couples to the right circular polarized state, and antiresonantly to the left circular polarized state, while the radially polarized highest energy state, on the contrary, resonantly couples to the left circular polarized state, and antiresonantly to the right circular polarized state. The change of the rotation direction will lead to the inversion of the coupling scheme, as it is shown in Fig.~\ref{FIG_Diags}(b).

In the rotating wave approximation the system of the four coupled equations thus splits into the two independent pairs, each of which coincides with well known equations for the description of the magnetic resonance of a spin, 
\begin{subequations}
\begin{eqnarray}	
&&\partial_t A_{T}=-i\omega_c(1-\Delta_1) A_T+i \eta e^{i \Omega t} A_{\uparrow},\\
&&\partial_t A_{\uparrow}=-i\omega_c A_{\uparrow}+i \eta e^{-i \Omega t} A_{T},
\label{two_fld_appr_1}
\end{eqnarray}
\end{subequations}
and
\begin{subequations}
\begin{eqnarray}
&&\partial_t A_{R}=-i\omega_c(1+\Delta_1) A_R-i \eta e^{-i \Omega t} A_{\downarrow}, \\
&&\partial_t A_{\downarrow}=-i\omega_c  A_{\downarrow}-i \eta e^{i \Omega t}A_{R}.
\label{two_fld_appr_2}
\end{eqnarray}
\end{subequations}

The application of the resonant rotating perturbation will thus lead to linear-circular polarization beats.  Note, that for a stationary potential ($\Omega=0$), two circular polarized components will be coupled instead~\cite{SupplMatRef}.

In resonant approximation one can easily get the simple analytical expression for the polarization occupancies.  For example, if at $t=0$ the lowest energy linear polarized state $A_T$ is populated, the occupancy of the circular polarized state resonantly coupled to it is
\begin{equation}
|A_\uparrow(t)|^2|=\frac{4\eta^2}{4\eta^2+\delta^2}\sin^2\left( \sqrt{\frac{\delta^2}{4}+\eta^2} t \right),     \label{occupation_dynamics}
\end{equation}
where $\delta=\Omega-\omega_c\Delta_1$ is the detuning from the exact resonance~\cite{SupplMatRef}.

\begin{figure}[tb!]
\begin{center}
 \includegraphics[width=\linewidth]{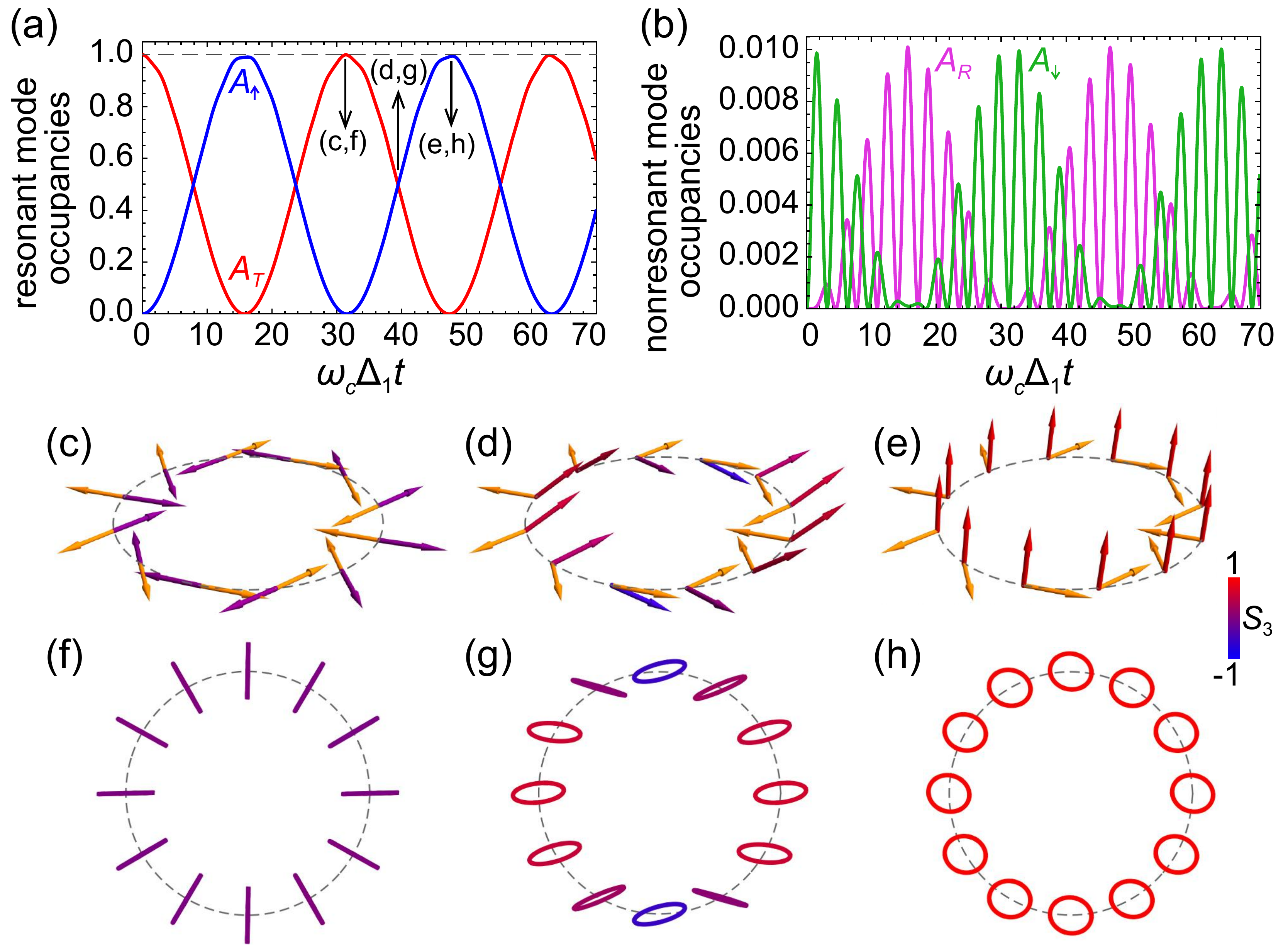}
\end{center}
\caption{ \label{figStokes_res}
(Color online) The dependencies of the occupancies of the resonant $A_{T,\uparrow}$ (a) and nonresonant $A_{R,\downarrow}$ (b) modes on time for the case of the exact resonance, $\omega_c \Delta_1 = \Omega$. 
The normalized Stokes vector $\vec{S}$ and the effective magnetic field at different angular positions on the ring are shown in panels (c)--(e) for the times indicated in (a).  The arrows  showing the Stokes vector are shown in color ranging from red to blue. 
Orange arrows indicate the orientation of the effective magnetic field. The polarization ellipses are shown in panels (f)--(h).
The parameters are ${\omega_c \Delta_1 = \Omega = 1}$,~${\eta=0.1}$.
}
\end{figure}

The temporal evolution of the resonant $A_{T,\uparrow} (t)$ and nonresonant $A_{R,\downarrow} (t)$ modes calculated from~\eqref{eq:coupled_modes_A} is illustrated in Figs.~\ref{figStokes_res}(a) and \ref{figStokes_res}(b), respectively.
One can see that the contribution of the nonresonant modes is negligibly small, and the formula~\eqref{occupation_dynamics} gives almost perfect approximation of the systems dynamics.

The normalized Stokes vectors $\vec{S} = (\psi ^{\dagger} \vec{\sigma} \psi ) / \psi ^{\dagger}\psi $ accompanied by polariztion ellipses at different spatial points are shown in panels (c)--(e) and  (f)--(h), respectively, for the times corresponding to the maximum occupancy of $A_T$ states, the maximum occupancy of $A_{\uparrow}$ state and the moment when these modes have the same occupancies.
$\vec{\sigma} = (\sigma_1 , \sigma_2 ,\sigma_3)$ is the vector of Pauli matrices.
One can clearly see the beatings between linear and circular polarized states, going through elliptically polarized states at intermediate times.

\textit{Floquet spectrum}. 
The coefficients in the system (\ref{eq:coupled_modes_A}) are periodic functions of time, and, according to the Floquet theorem the solutions of the system can be resperentes in the form
\begin{equation}
 A_j(t)=e^{-i\varepsilon t/\hbar}a_j(t),  
\end{equation}
where $a_j(t)=a_j(t+2\pi/\Omega)$ are periodic functions of time. The parameters $\varepsilon$ correspond to the so-called Floquet quasienergies of the system \cite{Shirley1965}. 

In our case, making the substitution $(a_1, a_2, a_2, a_4)^{\text{T}}= (b_1, b_2, b_3 e^{-i\Omega t}, b_4 e^{i\Omega t})^{\text{T}}$  we get:
\begin{eqnarray}	
i\frac{\partial\vec b}{\partial t} = \hat L \vec b, \label{eq:Floquet_spectrum}
\end{eqnarray}
where the matrix $\hat{L}$ is time independent,
\begin{eqnarray}	
\label{eq:Floquet_spectrum_operator}
\hat L=\left(\begin{array}{cccc} \omega_c(1-\Delta_1) & 0 & - \eta & - \eta \\ 
0 & \omega_c(1+\Delta_1) & - \eta &  \eta \\ 
-\eta & -\eta & \omega_c-\Omega &  0 \\
-\eta & \eta & 0 &  \omega_c + \Omega \\
\end{array}\right). \,\,
\end{eqnarray}
and Floquet quasienergies, up to a Planck constant, can be thus found as its eigenvalues. Thus we conclude that the problem of polariton states in a rotating potential can be conveniently considered in terms of Floquet states. 

\begin{figure}[tb!]
\begin{center}
\includegraphics[width=\linewidth]{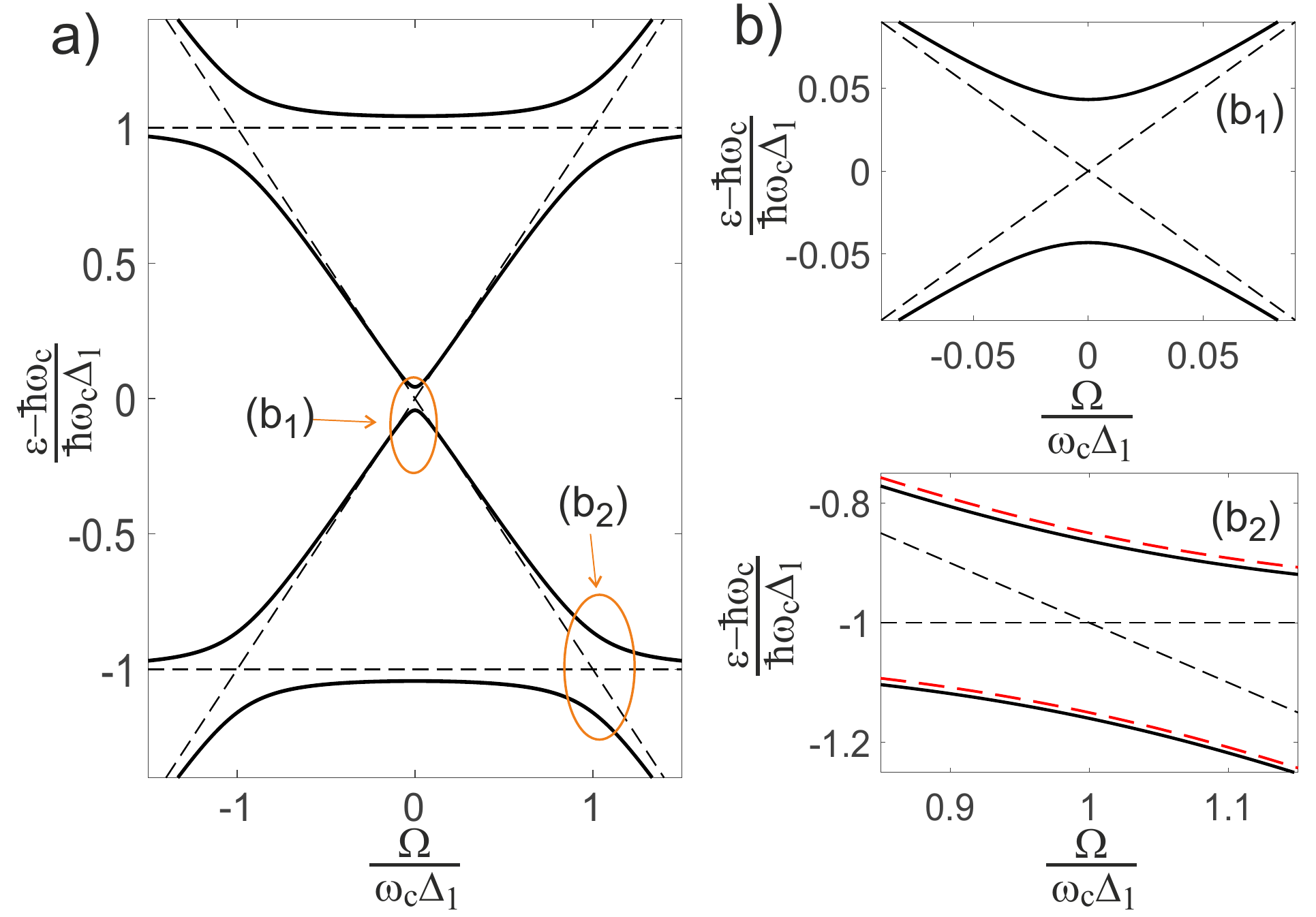}
\end{center}
\caption{ \label{Floquet_levels}
(Color online)  The dependencies of the Floquet eigenenergies on the potential rotation velocity $\Omega$ are shown in panel (a) for $\frac{ \eta }{ \omega_c \Delta_1 }=0.15$. The thin dashed lines correspond to the case $\eta=0 $. The panels (b$_1$) and (b$_2$) show the energy level splitting in the areas shown by the orange ovals in panel (a). The red lines in the right inset correspond to the eigenenergies calculated neglecting the anti-resonant terms. }
\end{figure}

Note that Floquet quasienergies can be found analytically, but corresponding expressions are bulky and they are not shown here. Instead we plot in Figure~\ref{Floquet_levels} the calculated dependencies of the Floquet energies on the rotation velocity~$\Omega$.  The presence of the rotating potential leads to the visible anticrossings of the Floquet quasienergies at $\Omega=0$ and $\Omega=\pm\omega_c \Delta_1$. Around $\Omega=0$ the anticrossing comes from the coupling between the states of the opposite circular polarizations $\psi_\uparrow$ and $\psi_\downarrow$, while at $\Omega=\pm\omega_c\Delta$ -- from the coupling between linear and circular polarized states, as it is shown in Fig.~\ref{FIG_Diags}.

Close to the resonance, in the rotating wave approximation Floquet quasienergies can be approximated as ~\cite{SupplMatRef}:
\begin{subequations}
\begin{eqnarray}
&&\varepsilon=\hbar \left( \omega_c + \frac{\omega_c \Delta_1 +\Omega}{2} \pm \sqrt{\frac{(\omega_c \Delta_1 -\Omega)^2}{4}+\eta^2} \right), \qquad \,\\
&&\varepsilon=\hbar \left( \omega_c - \frac{\omega_c \Delta_1 +\Omega}{2} \pm \sqrt{\frac{(\omega_c \Delta_1 -\Omega)^2}{4}+\eta^2} \right). \qquad \,
\end{eqnarray}
\end{subequations}
They are shown by the red lines in panel (b$_2$).  One can see that for a relatively shallow rotating potential the perturbation theory provides a very accurate estimate for  the eigenenergies. Note here that the limit of shallow potential is highly relevant to the experiments with optically induced rotating traps ~\cite{lagoudakis2022}.

\textit{Conclusions.}
In conclusion, we predict a parametric resonance leading to the polarization beats in polariton ring condensates subjected to a rotating perturbation. The considered effect is a polaritonic analogue of the electronic magnetic resonance. We demonstrate, that a rotating perturbation leads to the beats between linear and circular polarizations that manifest a cyclic dynamics of the polariton pseudospin. The phenomenon is a remarkable manifestation of the effect of mechanical rotation on spin properties of a quantum object. It may be used as a tool of control over the quantum state of a ring-shape polariton condensate which is important for applications in quantum and classical polariton computing.  Finally, let us notice that the incoherent pumping of polariton condensates creates complex effective potentials in general. The imaginary part of such a potential accounting for the interplay between losses and gain in each specific point of the real space. This interplay may lead to the pseudodrag effect ~\cite{Chestnov2019} that remained beyond the scope of the present study.

A.V.Y. and I.A.S. acknowledge financial support from Icelandic Research Fund (Rannis, the project ``Hybrid polaritonics''), ``Priority 2030 Academic Leadership Program'' and ``Goszadanie no. 2019-1246''.
A.V.K. and E.S.S. acknowledge Saint-Petersburg State University for the financial support (research grant No 91182694).
E.S.S. acknowledges support from the RF Ministry of Science and Higher Education as part of State Task no. 0635-2020-0013. 

 \bibliography{main_text_yulin}

\begin{thebibliography}{26}
\expandafter\ifx\csname natexlab\endcsname\relax\def\natexlab#1{#1}\fi
\expandafter\ifx\csname bibnamefont\endcsname\relax
  \def\bibnamefont#1{#1}\fi
\expandafter\ifx\csname bibfnamefont\endcsname\relax
  \def\bibfnamefont#1{#1}\fi
\expandafter\ifx\csname citenamefont\endcsname\relax
  \def\citenamefont#1{#1}\fi
\expandafter\ifx\csname url\endcsname\relax
  \def\url#1{\texttt{#1}}\fi
\expandafter\ifx\csname urlprefix\endcsname\relax\def\urlprefix{URL }\fi
\providecommand{\bibinfo}[2]{#2}
\providecommand{\eprint}[2][]{\url{#2}}

\bibitem[{\citenamefont{Ballarini et~al.}(2017)\citenamefont{Ballarini, Caputo,
  Mu\~noz, De~Giorgi, Dominici, Szyma\ifmmode~\acute{n}\else \'{n}\fi{}ska,
  West, Pfeiffer, Gigli, Laussy et~al.}}]{Ballarini2017}
\bibinfo{author}{\bibfnamefont{D.}~\bibnamefont{Ballarini}},
  \bibinfo{author}{\bibfnamefont{D.}~\bibnamefont{Caputo}},
  \bibinfo{author}{\bibfnamefont{C.~S.} \bibnamefont{Mu\~noz}},
  \bibinfo{author}{\bibfnamefont{M.}~\bibnamefont{De~Giorgi}},
  \bibinfo{author}{\bibfnamefont{L.}~\bibnamefont{Dominici}},
  \bibinfo{author}{\bibfnamefont{M.~H.}
  \bibnamefont{Szyma\ifmmode~\acute{n}\else \'{n}\fi{}ska}},
  \bibinfo{author}{\bibfnamefont{K.}~\bibnamefont{West}},
  \bibinfo{author}{\bibfnamefont{L.~N.} \bibnamefont{Pfeiffer}},
  \bibinfo{author}{\bibfnamefont{G.}~\bibnamefont{Gigli}},
  \bibinfo{author}{\bibfnamefont{F.~P.} \bibnamefont{Laussy}},
  \bibnamefont{et~al.}, \bibinfo{journal}{Phys. Rev. Lett.}
  \textbf{\bibinfo{volume}{118}}, \bibinfo{pages}{215301}
  (\bibinfo{year}{2017}),
  \urlprefix\url{https://link.aps.org/doi/10.1103/PhysRevLett.118.215301}.

\bibitem[{\citenamefont{Glazov et~al.}(2009)\citenamefont{Glazov, Ouerdane,
  Pilozzi, Malpuech, Kavokin, and D'Andrea}}]{Glazov2009}
\bibinfo{author}{\bibfnamefont{M.~M.} \bibnamefont{Glazov}},
  \bibinfo{author}{\bibfnamefont{H.}~\bibnamefont{Ouerdane}},
  \bibinfo{author}{\bibfnamefont{L.}~\bibnamefont{Pilozzi}},
  \bibinfo{author}{\bibfnamefont{G.}~\bibnamefont{Malpuech}},
  \bibinfo{author}{\bibfnamefont{A.~V.} \bibnamefont{Kavokin}},
  \bibnamefont{and} \bibinfo{author}{\bibfnamefont{A.}~\bibnamefont{D'Andrea}},
  \bibinfo{journal}{Phys. Rev. B} \textbf{\bibinfo{volume}{80}},
  \bibinfo{pages}{155306} (\bibinfo{year}{2009}),
  \urlprefix\url{https://link.aps.org/doi/10.1103/PhysRevB.80.155306}.

\bibitem[{\citenamefont{Bajoni et~al.}(2008)\citenamefont{Bajoni, Senellart,
  Wertz, Sagnes, Miard, Lema\^{\i}tre, and Bloch}}]{Bajoni2008}
\bibinfo{author}{\bibfnamefont{D.}~\bibnamefont{Bajoni}},
  \bibinfo{author}{\bibfnamefont{P.}~\bibnamefont{Senellart}},
  \bibinfo{author}{\bibfnamefont{E.}~\bibnamefont{Wertz}},
  \bibinfo{author}{\bibfnamefont{I.}~\bibnamefont{Sagnes}},
  \bibinfo{author}{\bibfnamefont{A.}~\bibnamefont{Miard}},
  \bibinfo{author}{\bibfnamefont{A.}~\bibnamefont{Lema\^{\i}tre}},
  \bibnamefont{and} \bibinfo{author}{\bibfnamefont{J.}~\bibnamefont{Bloch}},
  \bibinfo{journal}{Phys. Rev. Lett.} \textbf{\bibinfo{volume}{100}},
  \bibinfo{pages}{047401} (\bibinfo{year}{2008}),
  \urlprefix\url{https://link.aps.org/doi/10.1103/PhysRevLett.100.047401}.

\bibitem[{\citenamefont{Ctistis et~al.}(2010)\citenamefont{Ctistis, Hartsuiker,
  van~der Pol, Claudon, Vos, and G\'erard}}]{Ctistis2010}
\bibinfo{author}{\bibfnamefont{G.}~\bibnamefont{Ctistis}},
  \bibinfo{author}{\bibfnamefont{A.}~\bibnamefont{Hartsuiker}},
  \bibinfo{author}{\bibfnamefont{E.}~\bibnamefont{van~der Pol}},
  \bibinfo{author}{\bibfnamefont{J.}~\bibnamefont{Claudon}},
  \bibinfo{author}{\bibfnamefont{W.~L.} \bibnamefont{Vos}}, \bibnamefont{and}
  \bibinfo{author}{\bibfnamefont{J.-M.} \bibnamefont{G\'erard}},
  \bibinfo{journal}{Phys. Rev. B} \textbf{\bibinfo{volume}{82}},
  \bibinfo{pages}{195330} (\bibinfo{year}{2010}),
  \urlprefix\url{https://link.aps.org/doi/10.1103/PhysRevB.82.195330}.

\bibitem[{\citenamefont{Ferrier et~al.}(2011)\citenamefont{Ferrier, Wertz,
  Johne, Solnyshkov, Senellart, Sagnes, Lema\^{\i}tre, Malpuech, and
  Bloch}}]{Ferrier2011}
\bibinfo{author}{\bibfnamefont{L.}~\bibnamefont{Ferrier}},
  \bibinfo{author}{\bibfnamefont{E.}~\bibnamefont{Wertz}},
  \bibinfo{author}{\bibfnamefont{R.}~\bibnamefont{Johne}},
  \bibinfo{author}{\bibfnamefont{D.~D.} \bibnamefont{Solnyshkov}},
  \bibinfo{author}{\bibfnamefont{P.}~\bibnamefont{Senellart}},
  \bibinfo{author}{\bibfnamefont{I.}~\bibnamefont{Sagnes}},
  \bibinfo{author}{\bibfnamefont{A.}~\bibnamefont{Lema\^{\i}tre}},
  \bibinfo{author}{\bibfnamefont{G.}~\bibnamefont{Malpuech}}, \bibnamefont{and}
  \bibinfo{author}{\bibfnamefont{J.}~\bibnamefont{Bloch}},
  \bibinfo{journal}{Phys. Rev. Lett.} \textbf{\bibinfo{volume}{106}},
  \bibinfo{pages}{126401} (\bibinfo{year}{2011}),
  \urlprefix\url{https://link.aps.org/doi/10.1103/PhysRevLett.106.126401}.

\bibitem[{\citenamefont{Real et~al.}(2021)\citenamefont{Real, Carlon~Zambon,
  St-Jean, Sagnes, Lema\^{\i}tre, Le~Gratiet, Harouri, Ravets, Bloch, and
  Amo}}]{Real2021}
\bibinfo{author}{\bibfnamefont{B.}~\bibnamefont{Real}},
  \bibinfo{author}{\bibfnamefont{N.}~\bibnamefont{Carlon~Zambon}},
  \bibinfo{author}{\bibfnamefont{P.}~\bibnamefont{St-Jean}},
  \bibinfo{author}{\bibfnamefont{I.}~\bibnamefont{Sagnes}},
  \bibinfo{author}{\bibfnamefont{A.}~\bibnamefont{Lema\^{\i}tre}},
  \bibinfo{author}{\bibfnamefont{L.}~\bibnamefont{Le~Gratiet}},
  \bibinfo{author}{\bibfnamefont{A.}~\bibnamefont{Harouri}},
  \bibinfo{author}{\bibfnamefont{S.}~\bibnamefont{Ravets}},
  \bibinfo{author}{\bibfnamefont{J.}~\bibnamefont{Bloch}}, \bibnamefont{and}
  \bibinfo{author}{\bibfnamefont{A.}~\bibnamefont{Amo}},
  \bibinfo{journal}{Phys. Rev. Research} \textbf{\bibinfo{volume}{3}},
  \bibinfo{pages}{043161} (\bibinfo{year}{2021}),
  \urlprefix\url{https://link.aps.org/doi/10.1103/PhysRevResearch.3.043161}.

\bibitem[{\citenamefont{Galbiati et~al.}(2012)\citenamefont{Galbiati, Ferrier,
  Solnyshkov, Tanese, Wertz, Amo, Abbarchi, Senellart, Sagnes, Lema\^{\i}tre
  et~al.}}]{Galbiati2012}
\bibinfo{author}{\bibfnamefont{M.}~\bibnamefont{Galbiati}},
  \bibinfo{author}{\bibfnamefont{L.}~\bibnamefont{Ferrier}},
  \bibinfo{author}{\bibfnamefont{D.~D.} \bibnamefont{Solnyshkov}},
  \bibinfo{author}{\bibfnamefont{D.}~\bibnamefont{Tanese}},
  \bibinfo{author}{\bibfnamefont{E.}~\bibnamefont{Wertz}},
  \bibinfo{author}{\bibfnamefont{A.}~\bibnamefont{Amo}},
  \bibinfo{author}{\bibfnamefont{M.}~\bibnamefont{Abbarchi}},
  \bibinfo{author}{\bibfnamefont{P.}~\bibnamefont{Senellart}},
  \bibinfo{author}{\bibfnamefont{I.}~\bibnamefont{Sagnes}},
  \bibinfo{author}{\bibfnamefont{A.}~\bibnamefont{Lema\^{\i}tre}},
  \bibnamefont{et~al.}, \bibinfo{journal}{Phys. Rev. Lett.}
  \textbf{\bibinfo{volume}{108}}, \bibinfo{pages}{126403}
  (\bibinfo{year}{2012}),
  \urlprefix\url{https://link.aps.org/doi/10.1103/PhysRevLett.108.126403}.

\bibitem[{\citenamefont{Sala et~al.}(2015)\citenamefont{Sala, Solnyshkov,
  Carusotto, Jacqmin, Lema\^{\i}tre, Ter\ifmmode~\mbox{\c{c}}\else
  \c{c}\fi{}as, Nalitov, Abbarchi, Galopin, Sagnes et~al.}}]{Sala2015}
\bibinfo{author}{\bibfnamefont{V.~G.} \bibnamefont{Sala}},
  \bibinfo{author}{\bibfnamefont{D.~D.} \bibnamefont{Solnyshkov}},
  \bibinfo{author}{\bibfnamefont{I.}~\bibnamefont{Carusotto}},
  \bibinfo{author}{\bibfnamefont{T.}~\bibnamefont{Jacqmin}},
  \bibinfo{author}{\bibfnamefont{A.}~\bibnamefont{Lema\^{\i}tre}},
  \bibinfo{author}{\bibfnamefont{H.}~\bibnamefont{Ter\ifmmode~\mbox{\c{c}}\else
  \c{c}\fi{}as}}, \bibinfo{author}{\bibfnamefont{A.}~\bibnamefont{Nalitov}},
  \bibinfo{author}{\bibfnamefont{M.}~\bibnamefont{Abbarchi}},
  \bibinfo{author}{\bibfnamefont{E.}~\bibnamefont{Galopin}},
  \bibinfo{author}{\bibfnamefont{I.}~\bibnamefont{Sagnes}},
  \bibnamefont{et~al.}, \bibinfo{journal}{Phys. Rev. X}
  \textbf{\bibinfo{volume}{5}}, \bibinfo{pages}{011034} (\bibinfo{year}{2015}),
  \urlprefix\url{https://link.aps.org/doi/10.1103/PhysRevX.5.011034}.

\bibitem[{\citenamefont{Mili\ifmmode \acute{c}\else
  \'{c}\fi{}evi\ifmmode~\acute{c}\else \'{c}\fi{}
  et~al.}(2017)\citenamefont{Mili\ifmmode \acute{c}\else
  \'{c}\fi{}evi\ifmmode~\acute{c}\else \'{c}\fi{}, Ozawa, Montambaux,
  Carusotto, Galopin, Lema\^{\i}tre, Le~Gratiet, Sagnes, Bloch, and
  Amo}}]{Milicevic2017}
\bibinfo{author}{\bibfnamefont{M.}~\bibnamefont{Mili\ifmmode \acute{c}\else
  \'{c}\fi{}evi\ifmmode~\acute{c}\else \'{c}\fi{}}},
  \bibinfo{author}{\bibfnamefont{T.}~\bibnamefont{Ozawa}},
  \bibinfo{author}{\bibfnamefont{G.}~\bibnamefont{Montambaux}},
  \bibinfo{author}{\bibfnamefont{I.}~\bibnamefont{Carusotto}},
  \bibinfo{author}{\bibfnamefont{E.}~\bibnamefont{Galopin}},
  \bibinfo{author}{\bibfnamefont{A.}~\bibnamefont{Lema\^{\i}tre}},
  \bibinfo{author}{\bibfnamefont{L.}~\bibnamefont{Le~Gratiet}},
  \bibinfo{author}{\bibfnamefont{I.}~\bibnamefont{Sagnes}},
  \bibinfo{author}{\bibfnamefont{J.}~\bibnamefont{Bloch}}, \bibnamefont{and}
  \bibinfo{author}{\bibfnamefont{A.}~\bibnamefont{Amo}},
  \bibinfo{journal}{Phys. Rev. Lett.} \textbf{\bibinfo{volume}{118}},
  \bibinfo{pages}{107403} (\bibinfo{year}{2017}),
  \urlprefix\url{https://link.aps.org/doi/10.1103/PhysRevLett.118.107403}.

\bibitem[{\citenamefont{Suchomel et~al.}(2018)\citenamefont{Suchomel, Klembt,
  Harder, Klaas, Egorov, Winkler, Emmerling, Thomale, H\"ofling, and
  Schneider}}]{Suchomel2018}
\bibinfo{author}{\bibfnamefont{H.}~\bibnamefont{Suchomel}},
  \bibinfo{author}{\bibfnamefont{S.}~\bibnamefont{Klembt}},
  \bibinfo{author}{\bibfnamefont{T.~H.} \bibnamefont{Harder}},
  \bibinfo{author}{\bibfnamefont{M.}~\bibnamefont{Klaas}},
  \bibinfo{author}{\bibfnamefont{O.~A.} \bibnamefont{Egorov}},
  \bibinfo{author}{\bibfnamefont{K.}~\bibnamefont{Winkler}},
  \bibinfo{author}{\bibfnamefont{M.}~\bibnamefont{Emmerling}},
  \bibinfo{author}{\bibfnamefont{R.}~\bibnamefont{Thomale}},
  \bibinfo{author}{\bibfnamefont{S.}~\bibnamefont{H\"ofling}},
  \bibnamefont{and}
  \bibinfo{author}{\bibfnamefont{C.}~\bibnamefont{Schneider}},
  \bibinfo{journal}{Phys. Rev. Lett.} \textbf{\bibinfo{volume}{121}},
  \bibinfo{pages}{257402} (\bibinfo{year}{2018}),
  \urlprefix\url{https://link.aps.org/doi/10.1103/PhysRevLett.121.257402}.

\bibitem[{\citenamefont{Whittaker et~al.}(2018)\citenamefont{Whittaker,
  Cancellieri, Walker, Gulevich, Schomerus, Vaitiekus, Royall, Whittaker,
  Clarke, Iorsh et~al.}}]{Whittaker2018}
\bibinfo{author}{\bibfnamefont{C.~E.} \bibnamefont{Whittaker}},
  \bibinfo{author}{\bibfnamefont{E.}~\bibnamefont{Cancellieri}},
  \bibinfo{author}{\bibfnamefont{P.~M.} \bibnamefont{Walker}},
  \bibinfo{author}{\bibfnamefont{D.~R.} \bibnamefont{Gulevich}},
  \bibinfo{author}{\bibfnamefont{H.}~\bibnamefont{Schomerus}},
  \bibinfo{author}{\bibfnamefont{D.}~\bibnamefont{Vaitiekus}},
  \bibinfo{author}{\bibfnamefont{B.}~\bibnamefont{Royall}},
  \bibinfo{author}{\bibfnamefont{D.~M.} \bibnamefont{Whittaker}},
  \bibinfo{author}{\bibfnamefont{E.}~\bibnamefont{Clarke}},
  \bibinfo{author}{\bibfnamefont{I.~V.} \bibnamefont{Iorsh}},
  \bibnamefont{et~al.}, \bibinfo{journal}{Phys. Rev. Lett.}
  \textbf{\bibinfo{volume}{120}}, \bibinfo{pages}{097401}
  (\bibinfo{year}{2018}),
  \urlprefix\url{https://link.aps.org/doi/10.1103/PhysRevLett.120.097401}.

\bibitem[{\citenamefont{Whittaker et~al.}(2021)\citenamefont{Whittaker,
  Dowling, Nalitov, Yulin, Royall, Clarke, Skolnick, Shelykh, and
  Krizhanovskii}}]{Whittaker2021}
\bibinfo{author}{\bibfnamefont{C.~E.} \bibnamefont{Whittaker}},
  \bibinfo{author}{\bibfnamefont{T.}~\bibnamefont{Dowling}},
  \bibinfo{author}{\bibfnamefont{A.~V.} \bibnamefont{Nalitov}},
  \bibinfo{author}{\bibfnamefont{A.~V.} \bibnamefont{Yulin}},
  \bibinfo{author}{\bibfnamefont{B.}~\bibnamefont{Royall}},
  \bibinfo{author}{\bibfnamefont{E.}~\bibnamefont{Clarke}},
  \bibinfo{author}{\bibfnamefont{M.~S.} \bibnamefont{Skolnick}},
  \bibinfo{author}{\bibfnamefont{I.~A.} \bibnamefont{Shelykh}},
  \bibnamefont{and} \bibinfo{author}{\bibfnamefont{D.~N.}
  \bibnamefont{Krizhanovskii}}, \bibinfo{journal}{Nat. Photon.}
  \textbf{\bibinfo{volume}{15}}, \bibinfo{pages}{193} (\bibinfo{year}{2021}),
  \urlprefix\url{https://www.nature.com/articles/s41566-020-00729-z}.

\bibitem[{\citenamefont{Kuriakose et~al.}(2022)\citenamefont{Kuriakose, Walker,
  Dowling, Kyriienko, Shelykh, St-Jean, Carlon~Zambon, Lemaitre, Sagnes,
  Legratiet et~al.}}]{Kuriakose2022}
\bibinfo{author}{\bibfnamefont{T.}~\bibnamefont{Kuriakose}},
  \bibinfo{author}{\bibfnamefont{P.~M.} \bibnamefont{Walker}},
  \bibinfo{author}{\bibfnamefont{T.}~\bibnamefont{Dowling}},
  \bibinfo{author}{\bibfnamefont{O.}~\bibnamefont{Kyriienko}},
  \bibinfo{author}{\bibfnamefont{I.~A.} \bibnamefont{Shelykh}},
  \bibinfo{author}{\bibfnamefont{P.}~\bibnamefont{St-Jean}},
  \bibinfo{author}{\bibfnamefont{N.}~\bibnamefont{Carlon~Zambon}},
  \bibinfo{author}{\bibfnamefont{A.}~\bibnamefont{Lemaitre}},
  \bibinfo{author}{\bibfnamefont{I.}~\bibnamefont{Sagnes}},
  \bibinfo{author}{\bibfnamefont{L.}~\bibnamefont{Legratiet}},
  \bibnamefont{et~al.}, \bibinfo{journal}{Nat. Photon.}
  \textbf{\bibinfo{volume}{16}}, \bibinfo{pages}{566} (\bibinfo{year}{2022}),
  \urlprefix\url{https://www.nature.com/articles/s41566-022-01019-6}.

\bibitem[{\citenamefont{Lukoshkin et~al.}(2018)\citenamefont{Lukoshkin,
  Kalevich, Afanasiev, Kavokin, Hatzopoulos, Savvidis, Sedov, and
  Kavokin}}]{Lukoshkin2018}
\bibinfo{author}{\bibfnamefont{V.~A.} \bibnamefont{Lukoshkin}},
  \bibinfo{author}{\bibfnamefont{V.~K.} \bibnamefont{Kalevich}},
  \bibinfo{author}{\bibfnamefont{M.~M.} \bibnamefont{Afanasiev}},
  \bibinfo{author}{\bibfnamefont{K.~V.} \bibnamefont{Kavokin}},
  \bibinfo{author}{\bibfnamefont{Z.}~\bibnamefont{Hatzopoulos}},
  \bibinfo{author}{\bibfnamefont{P.~G.} \bibnamefont{Savvidis}},
  \bibinfo{author}{\bibfnamefont{E.~S.} \bibnamefont{Sedov}}, \bibnamefont{and}
  \bibinfo{author}{\bibfnamefont{A.~V.} \bibnamefont{Kavokin}},
  \bibinfo{journal}{Phys. Rev. B} \textbf{\bibinfo{volume}{97}},
  \bibinfo{pages}{195149} (\bibinfo{year}{2018}),
  \urlprefix\url{https://link.aps.org/doi/10.1103/PhysRevB.97.195149}.

\bibitem[{\citenamefont{Sedov et~al.}(2021{\natexlab{a}})\citenamefont{Sedov,
  Arakelian, and Kavokin}}]{SciRep1122382}
\bibinfo{author}{\bibfnamefont{E.}~\bibnamefont{Sedov}},
  \bibinfo{author}{\bibfnamefont{S.}~\bibnamefont{Arakelian}},
  \bibnamefont{and} \bibinfo{author}{\bibfnamefont{A.}~\bibnamefont{Kavokin}},
  \bibinfo{journal}{Scientific Reports} \textbf{\bibinfo{volume}{11}},
  \bibinfo{pages}{22382} (\bibinfo{year}{2021}{\natexlab{a}}),
  \urlprefix\url{https://doi.org/10.1038/s41598-021-01812-3}.

\bibitem[{\citenamefont{Shelykh et~al.}(2009)\citenamefont{Shelykh, Pavlovic,
  Solnyshkov, and Malpuech}}]{Shelykh2010}
\bibinfo{author}{\bibfnamefont{I.~A.} \bibnamefont{Shelykh}},
  \bibinfo{author}{\bibfnamefont{G.}~\bibnamefont{Pavlovic}},
  \bibinfo{author}{\bibfnamefont{D.~D.} \bibnamefont{Solnyshkov}},
  \bibnamefont{and} \bibinfo{author}{\bibfnamefont{G.}~\bibnamefont{Malpuech}},
  \bibinfo{journal}{Phys. Rev. Lett.} \textbf{\bibinfo{volume}{102}},
  \bibinfo{pages}{046407} (\bibinfo{year}{2009}),
  \urlprefix\url{https://link.aps.org/doi/10.1103/PhysRevLett.102.046407}.

\bibitem[{\citenamefont{Zezyulin et~al.}(2018)\citenamefont{Zezyulin, Gulevich,
  Skryabin, and Shelykh}}]{Zezyulin2018}
\bibinfo{author}{\bibfnamefont{D.~A.} \bibnamefont{Zezyulin}},
  \bibinfo{author}{\bibfnamefont{D.~R.} \bibnamefont{Gulevich}},
  \bibinfo{author}{\bibfnamefont{D.~V.} \bibnamefont{Skryabin}},
  \bibnamefont{and} \bibinfo{author}{\bibfnamefont{I.~A.}
  \bibnamefont{Shelykh}}, \bibinfo{journal}{Phys. Rev. B}
  \textbf{\bibinfo{volume}{97}}, \bibinfo{pages}{161302}
  (\bibinfo{year}{2018}),
  \urlprefix\url{https://link.aps.org/doi/10.1103/PhysRevB.97.161302}.

\bibitem[{\citenamefont{Gulevich et~al.}(2016)\citenamefont{Gulevich, Skryabin,
  Alodjants, and Shelykh}}]{Gulevich2016}
\bibinfo{author}{\bibfnamefont{D.~R.} \bibnamefont{Gulevich}},
  \bibinfo{author}{\bibfnamefont{D.~V.} \bibnamefont{Skryabin}},
  \bibinfo{author}{\bibfnamefont{A.~P.} \bibnamefont{Alodjants}},
  \bibnamefont{and} \bibinfo{author}{\bibfnamefont{I.~A.}
  \bibnamefont{Shelykh}}, \bibinfo{journal}{Phys. Rev. B}
  \textbf{\bibinfo{volume}{94}}, \bibinfo{pages}{115407}
  (\bibinfo{year}{2016}),
  \urlprefix\url{https://link.aps.org/doi/10.1103/PhysRevB.94.115407}.

\bibitem[{\citenamefont{Sedov et~al.}(2021{\natexlab{b}})\citenamefont{Sedov,
  Lukoshkin, Kalevich, Savvidis, and Kavokin}}]{Sedov2022}
\bibinfo{author}{\bibfnamefont{E.~S.} \bibnamefont{Sedov}},
  \bibinfo{author}{\bibfnamefont{V.~A.} \bibnamefont{Lukoshkin}},
  \bibinfo{author}{\bibfnamefont{V.~K.} \bibnamefont{Kalevich}},
  \bibinfo{author}{\bibfnamefont{P.~G.} \bibnamefont{Savvidis}},
  \bibnamefont{and} \bibinfo{author}{\bibfnamefont{A.~V.}
  \bibnamefont{Kavokin}}, \bibinfo{journal}{Phys. Rev. Research}
  \textbf{\bibinfo{volume}{3}}, \bibinfo{pages}{013072}
  (\bibinfo{year}{2021}{\natexlab{b}}),
  \urlprefix\url{https://link.aps.org/doi/10.1103/PhysRevResearch.3.013072}.

\bibitem[{\citenamefont{Xue et~al.}(2021)\citenamefont{Xue, Chestnov, Sedov,
  Kiktenko, Fedorov, Schumacher, Ma, and Kavokin}}]{Xue2021}
\bibinfo{author}{\bibfnamefont{Y.}~\bibnamefont{Xue}},
  \bibinfo{author}{\bibfnamefont{I.}~\bibnamefont{Chestnov}},
  \bibinfo{author}{\bibfnamefont{E.}~\bibnamefont{Sedov}},
  \bibinfo{author}{\bibfnamefont{E.}~\bibnamefont{Kiktenko}},
  \bibinfo{author}{\bibfnamefont{A.~K.} \bibnamefont{Fedorov}},
  \bibinfo{author}{\bibfnamefont{S.}~\bibnamefont{Schumacher}},
  \bibinfo{author}{\bibfnamefont{X.}~\bibnamefont{Ma}}, \bibnamefont{and}
  \bibinfo{author}{\bibfnamefont{A.}~\bibnamefont{Kavokin}},
  \bibinfo{journal}{Phys. Rev. Research} \textbf{\bibinfo{volume}{3}},
  \bibinfo{pages}{013099} (\bibinfo{year}{2021}),
  \urlprefix\url{https://link.aps.org/doi/10.1103/PhysRevResearch.3.013099}.

\bibitem[{\citenamefont{Gnusov et~al.}(2023)\citenamefont{Gnusov, Harrison,
  Alyatkin, Sitnik, T\"{o}pfer, Sigurdsson, and Lagoudakis}}]{lagoudakis2022}
\bibinfo{author}{\bibfnamefont{I.}~\bibnamefont{Gnusov}},
  \bibinfo{author}{\bibfnamefont{S.}~\bibnamefont{Harrison}},
  \bibinfo{author}{\bibfnamefont{S.}~\bibnamefont{Alyatkin}},
  \bibinfo{author}{\bibfnamefont{K.}~\bibnamefont{Sitnik}},
  \bibinfo{author}{\bibfnamefont{J.}~\bibnamefont{T\"{o}pfer}},
  \bibinfo{author}{\bibfnamefont{H.}~\bibnamefont{Sigurdsson}},
  \bibnamefont{and}
  \bibinfo{author}{\bibfnamefont{P.}~\bibnamefont{Lagoudakis}},
  \bibinfo{journal}{Science Advances} \textbf{\bibinfo{volume}{0}},
  \bibinfo{pages}{0} (\bibinfo{year}{2023}),
  \urlprefix\url{https://link.aps.org/doi/10.1103/PhysRevResearch.3.013099}.

\bibitem[{\citenamefont{Kozin et~al.}(2018)\citenamefont{Kozin, Shelykh,
  Nalitov, and Iorsh}}]{Kozin2018}
\bibinfo{author}{\bibfnamefont{V.~K.} \bibnamefont{Kozin}},
  \bibinfo{author}{\bibfnamefont{I.~A.} \bibnamefont{Shelykh}},
  \bibinfo{author}{\bibfnamefont{A.~V.} \bibnamefont{Nalitov}},
  \bibnamefont{and} \bibinfo{author}{\bibfnamefont{I.~V.} \bibnamefont{Iorsh}},
  \bibinfo{journal}{Phys. Rev. B} \textbf{\bibinfo{volume}{98}},
  \bibinfo{pages}{125115} (\bibinfo{year}{2018}),
  \urlprefix\url{https://link.aps.org/doi/10.1103/PhysRevB.98.125115}.

\bibitem[{\citenamefont{Rubo}(2022)}]{Rubo2022}
\bibinfo{author}{\bibfnamefont{Y.~G.} \bibnamefont{Rubo}},
  \bibinfo{journal}{Phys. Rev. B} \textbf{\bibinfo{volume}{106}},
  \bibinfo{pages}{235306} (\bibinfo{year}{2022}),
  \urlprefix\url{https://link.aps.org/doi/10.1103/PhysRevB.106.235306}.

\bibitem[{Sup()}]{SupplMatRef}
\bibinfo{note}{See Supplemental Material for the corrections to the evolution
  of the beating polarization modes die to the contribution of non-resonant
  terms, derivation of occupancies of the beating modes and analysis of the
  polarization beating in the case of the resting potential.}

\bibitem[{\citenamefont{Shirley}(1965)}]{Shirley1965}
\bibinfo{author}{\bibfnamefont{J.~H.} \bibnamefont{Shirley}},
  \bibinfo{journal}{Phys. Rev.} \textbf{\bibinfo{volume}{138}},
  \bibinfo{pages}{B979} (\bibinfo{year}{1965}),
  \urlprefix\url{https://link.aps.org/doi/10.1103/PhysRev.138.B979}.

\bibitem[{\citenamefont{Chestnov et~al.}(2019)\citenamefont{Chestnov, Rubo, and
  Kavokin}}]{Chestnov2019}
\bibinfo{author}{\bibfnamefont{I.~Y.} \bibnamefont{Chestnov}},
  \bibinfo{author}{\bibfnamefont{Y.~G.} \bibnamefont{Rubo}}, \bibnamefont{and}
  \bibinfo{author}{\bibfnamefont{A.~V.} \bibnamefont{Kavokin}},
  \bibinfo{journal}{Phys. Rev. B} \textbf{\bibinfo{volume}{100}},
  \bibinfo{pages}{085302} (\bibinfo{year}{2019}),
  \urlprefix\url{https://link.aps.org/doi/10.1103/PhysRevB.100.85302}.

\end{thebibliography}

\end{document}


\title{Supplementary Material: Spin resonance induced by a mechanical rotation of a polariton condensate}

\author{A.V. Yulin}
\affiliation{Department of Physics, ITMO University, Saint Petersburg 197101, Russia}
\affiliation{Science Institute, University of Iceland, Dunhagi 3, IS-107, Reykjavik, Iceland}

\author{I.A. Shelykh}
\affiliation{Science Institute, University of Iceland, Dunhagi 3, IS-107, Reykjavik, Iceland}
\affiliation{Department of Physics, ITMO University, Saint Petersburg 197101, Russia}

\author{E. S. Sedov}
\affiliation{Russian Quantum Center, Skolkovo, Moscow 143025, Russia}
\affiliation{Spin-Optics laboratory, St. Petersburg State University, St. Petersburg 198504, Russia}
\affiliation{Vladimir State University, Vladimir 600000, Russia}

\author{A.V. Kavokin}
\affiliation{Westlake University, School of Science, 18 Shilongshan Road, Hangzhou 310024, Zhejiang Province, China}
\affiliation{Westlake Institute for Advanced Study, Institute of Natural Sciences, 18 Shilongshan Road, Hangzhou 310024, Zhejiang Province, China}
\affiliation{Spin-Optics laboratory, St. Petersburg State University, St. Petersburg 198504, Russia}

\date{\today}

\maketitle

\section{Corrections to the coupling strength due to non-resonant terms}

To develop the perturbation theory it is convenient to write the equations in Floquet basis introduced in the main text.
\begin{subequations}
	\label{eq:Floquet_spectrum}
\begin{eqnarray}	
&&\frac{\varepsilon}{\hbar} b_{1}=\omega_c(1-\Delta_1) b_1 - \eta \left( b_{3} + b_{4}\right), \, \\
&&\frac{\varepsilon}{\hbar} b_{2}=\omega_c(1+\Delta_1) b_2 -  \eta \left(  b_{3} - b_{4}\right), \, \\
&&\frac{\varepsilon}{\hbar} b_{3}=(\omega_c -\Omega) b_{3} - \eta (b_{1}+b_{2}), \,  \\
&&\frac{\varepsilon}{\hbar} b_{4}=(\omega_c +\Omega) b_{4} - \eta (b_{1}-b_{2}). \,
\end{eqnarray}
\end{subequations}
Let us assume that $\frac{ |\Omega - \omega_c \Delta_1|}{|\Omega + \omega_c \Delta_1|} \ll 1$ and, for concreteness that $\frac{\varepsilon}{\hbar}-\omega_c>0$. If $|\eta | \ll |\omega_c \Delta_1| $ the energy splitting at the exact resonance is much less compared to TE-TM splitting $| \frac{\varepsilon}{\hbar} -\omega_c | \ll |\omega_c \Delta_1|$ and this allows to express $b_2$ and $b_4$ through $b_1$ and $b_3$ from (\ref{eq:Floquet_spectrum}b,d)
\begin{subequations}
\label{SEQb2andb4}
\begin{equation*}
b_2=\frac{\eta}{\omega_c \Delta_1}b_3, \qquad
b_4=\frac{\eta}{\omega_c \Delta_1}b_1.
\eqno{(\text{\ref{SEQb2andb4}a,b})}
\end{equation*}
\end{subequations}
Substituting~\eqref{SEQb2andb4} into (\ref{eq:Floquet_spectrum}a,c) we obtain
\begin{subequations}
\begin{eqnarray}
&&\frac{\varepsilon}{\hbar} b_{1}=\left( \omega_c-\omega_c\Delta_1- \frac{\eta^2}{\omega_c \Delta_1} \right) b_1 - \eta b_{3}, \, \\
&&\frac{\varepsilon}{\hbar} b_{3}=\left( \omega_c -\Omega-\frac{\eta^2}{\omega_c \Delta_1} \right) b_{3} - \eta b_{1}.   
\end{eqnarray}
\end{subequations}

From this we derive the expression for the eigenenergies 
\begin{eqnarray}
\varepsilon=\hbar \left[ \omega_c (1-\Delta_1) -\frac{\eta^2}{\omega_c \Delta_1}+ \omega_{\pm} \right], \label{eq:corrected_frq}
\end{eqnarray}
where
$\omega_{\pm}=-\frac{\delta}{2} \pm  \sqrt{\frac{\delta^2}{4}+\eta^2} $ and
$\delta= \Omega - \omega_c \Delta_1 $.
Thus we can conclude that in the first approximation order the non-resonant terms do not contribute to the splitting of the eigenfrequencies in the vicinity of $\Omega=\omega_c \Delta_1$. The only effect is that the eigenenergies get shifted by $\frac{\eta^2}{\omega_c \Delta_1}$. 
The analogues procedure can be done for $\frac{\varepsilon}{\hbar}-\omega_c < 0$. In the same way we can consider the resonances appearing at $\Omega \approx - \omega_c \Delta_1$. 

To show that the initial excitation in the form of a radially or tangentially polarized state will never transform to a pure circularly polarized state  we can return to the initial basis and write down the expression for the field keeping also the corrections appearing due to the non-resonant terms
\begin{widetext}
\begin{eqnarray}
\vec \psi_{\pm}=\left[ \frac{\eta}{\sqrt{2}}\left(\begin{array}{c} e^{-i\theta}+\frac{\omega_{\pm}}{\omega_c\Delta_1}e^{i\theta} \\ e^{i\theta}-\frac{\omega_{\pm}}{\omega_c\Delta_1} e^{-i\theta} \end{array}\right)+  \omega_{\pm}\left(\begin{array}{c}e^{i\theta} \\ 0 \end{array}\right) e^{-i\Omega t}   +  \frac{\eta^2}{\omega_c\Delta_1}\left(\begin{array}{c} 0 \\ e^{-i\theta} \end{array}\right) e^{i\Omega t}   \right] e^{-i[\omega_c(1-\Delta_1)+\frac{\eta^2}{2\omega_c \Delta_1}+\omega_{\pm}]t}.
\label{eigenstates_general}
\end{eqnarray}
\end{widetext}
It is seen that the radially or tangentionally polarized state cannot be represented as a composition of just two modes $\vec \psi_{\pm}$ but all four modes are required. The frequencies of the modes are not commensurable in a general case and thus the pure linearly polarized state will never reappear in the system. 

However if we disregard the non-resonant terms then the eigenmodes are 
\begin{widetext}
\begin{eqnarray}
\vec \psi_{\pm}=\left[ \frac{\eta}{\sqrt{2}}\left(\begin{array}{c} e^{-i\theta} \\ e^{i\theta} \end{array}\right) +\omega_{\pm}  \left(\begin{array}{c}e^{i\theta} \\ 0 \end{array}\right) e^{-i\Omega t}      \right] e^{-i[\omega_c(1-\Delta_1) +\omega_{\pm}]t}.
\label{eigenstates_simplf}
\end{eqnarray}
\end{widetext}
This means that the initial radially or tangentialy polarized state can be represents as a sum of just to modes and so it will be completely restored after evolution for $\Delta t=\frac{\pi}{\eta}$. 

\section{Polarizations beats for $\Omega \approx \omega_c \Delta_1$ }

Let us derive the formula describing the occupancy of the modes $A_{T}$ and $A_{\uparrow}$ for the case when the non-resonant terms can be neglected. In this case 
the solution can be found in the form $\vec \psi= C_{+} \vec \psi_{+} + C_{-} \vec \psi_{-}$ where the eigenmodes $\vec \psi_{\pm}$ are given by (\ref{eigenstates_simplf}).
Requiring that the initial state is tangential polarized we obtain $C_{-}= - \frac{\omega_{+}}{\omega_{-}}C_{+}$. Then from the condition that the initial occupancy $\rho=|\vec \psi|^2$ of the tangential polarized stated is equal to unity we obtain (up to an arbitrary phase that can be set to zero without loss of generality) that 
\begin{subequations}
\label{SEQCcoeffs}
\begin{equation*}
C_{+}=\frac{\omega_{-}}{\eta (\omega_{-}-\omega_{+} )}, \qquad 
C_{-}=\frac{\omega_{+}}{\eta (\omega_{+}-\omega_{-} )}.
\eqno{(\text{\ref{SEQCcoeffs}a,b})}
\end{equation*}
\end{subequations}

Using the expressions for $C_{\pm}$, after some algebra, one obtains  
\begin{widetext}
\begin{eqnarray}
\vec \psi =\left[ \frac{1}{\sqrt{2}}\frac{\omega_{+} e^{-i\omega_{-}t }-\omega_{-} e^{-i\omega_{+}t }}{\omega_{+}-\omega_{-}}\left(\begin{array}{c} e^{-i\theta} \\ e^{i\theta} \end{array}\right) -\frac{\omega_{+} \omega_{-}  }{\eta (\omega_{+}-\omega_{-})} ( e^{-i\omega_{+}t}-e^{-i\omega_{-}t} )  \left(\begin{array}{c}e^{i\theta} \\ 0 \end{array}\right) e^{-i\Omega t}      \right] e^{-i\omega_c(1-\Delta_1) t}.
\label{eigenstates_simpl_st_sp}
\end{eqnarray}
\end{widetext}
Therefore for the occupancies of the tangential $\rho_{T}$ and circularly $\rho_{\uparrow}$ states we have
\begin{subequations}
\label{eigenstates_simpl12}
\begin{eqnarray}
\label{eigenstates_simpl_occup1}
&&\rho_{T}=\frac{1}{\delta^2+4\eta^2} \nonumber \\
&& \hphantom{\rho_{T}} \times\left\{ \delta^2 
+ 4\eta^2 \left[1- \sin^2 \left(\sqrt{\frac{\delta^2}{4}+\eta^2}t \right) \right] \right\}, \quad \\
&&\rho_{\uparrow}=\frac{4\eta^2}{\delta^2+4\eta^2} \sin^2 \left[ \sqrt{\frac{\delta^2}{4}+\eta^2}t \right].  \, \,\,
\label{eigenstates_simpl_occup2}
\end{eqnarray}
\end{subequations}
As one should expect the beating period $T_b$ is equal to inverse difference of the eigenfrequencies of the modes $T_b= {2 \pi }/{|\omega_{+}-\omega_{-}|}={\pi}({\delta^2}/{4}+\eta^2)^{-1/2}$.

\begin{figure}[tb]
\begin{center}
 \includegraphics[width=\linewidth]{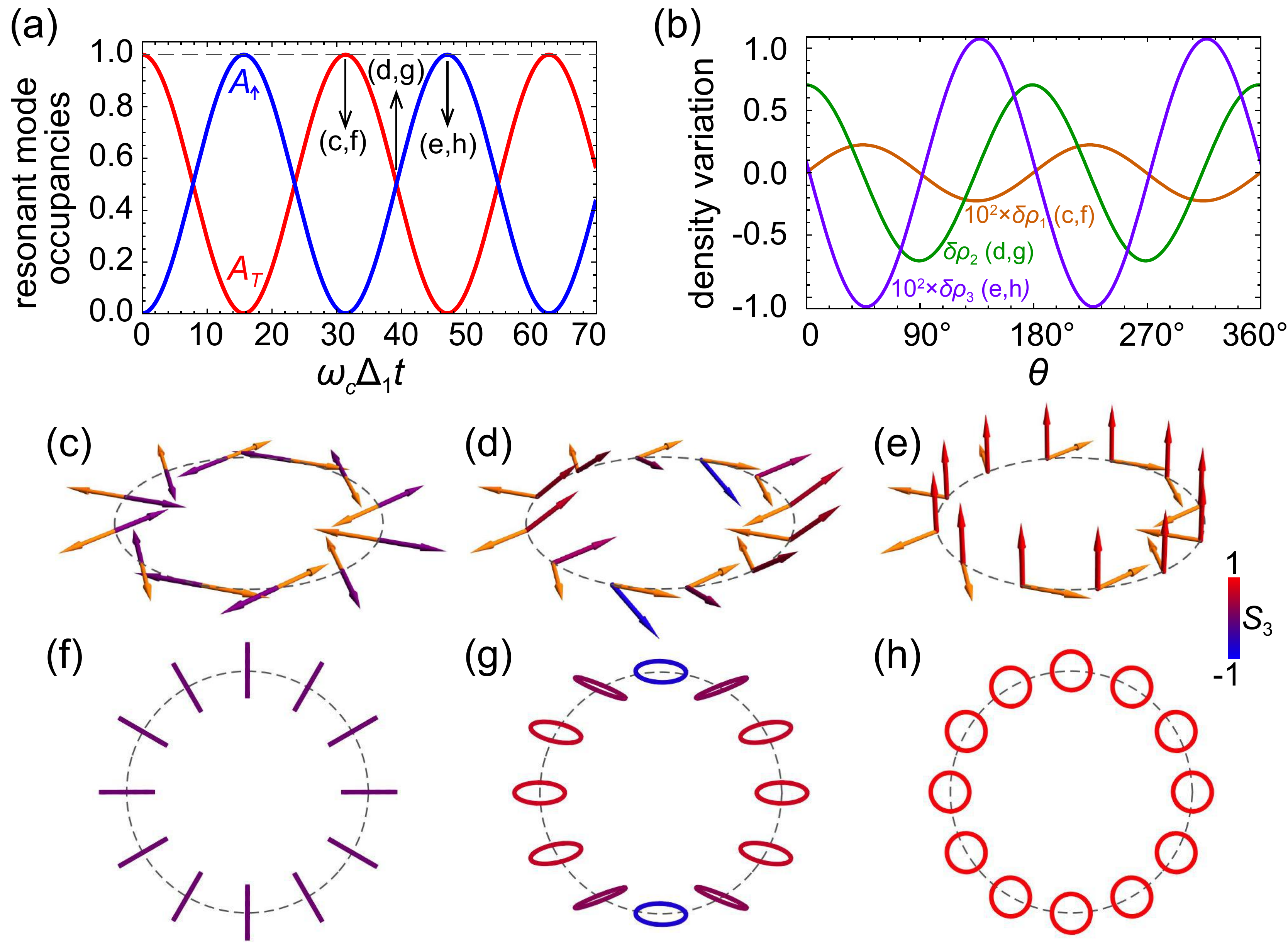}
\end{center}
\caption{ \label{SUPPL_figSphereRES01}
(Color online) The dependencies of the occupancies of the resonant $A_{T,\uparrow}$ modes on time for the case of the exact resonance, $\omega_c \Delta_1 = \Omega$, when we disregard the nonresonant terms in Eqs.~(7) in the main text (a).
Total density variation around the ring $\delta \rho = \left(|\vec{\psi}|^2 - 1 \right)$ for the times indicated in~(a).
Due to small deviation of the density from 1, the density variations $\delta \rho _{1,3}$ are shown with factor~$10^2$.
The normalized Stokes vector $\vec{S}$ and the effective magnetic field at different angular positions on the ring are shown in panels (c)--(e) for the times indicated in (a).  The arrows  showing the Stokes vector are shown in color ranging from red to blue. 
Orange arrows indicate the orientation of the effective magnetic field. The polarization ellipses are shown in panels (f)--(h).
The parameters are $\omega_c \Delta_1=1$, $\eta=0.1$.
The initial conditions vector is 
${[A_{\text{T}}(0),A_{\text{R}}(0),A_{\uparrow}(0),A_{\downarrow}(0)] = (1,0,0,0)}$.
}
\end{figure}

\begin{figure}[tb]
\begin{center}
 \includegraphics[width=\linewidth]{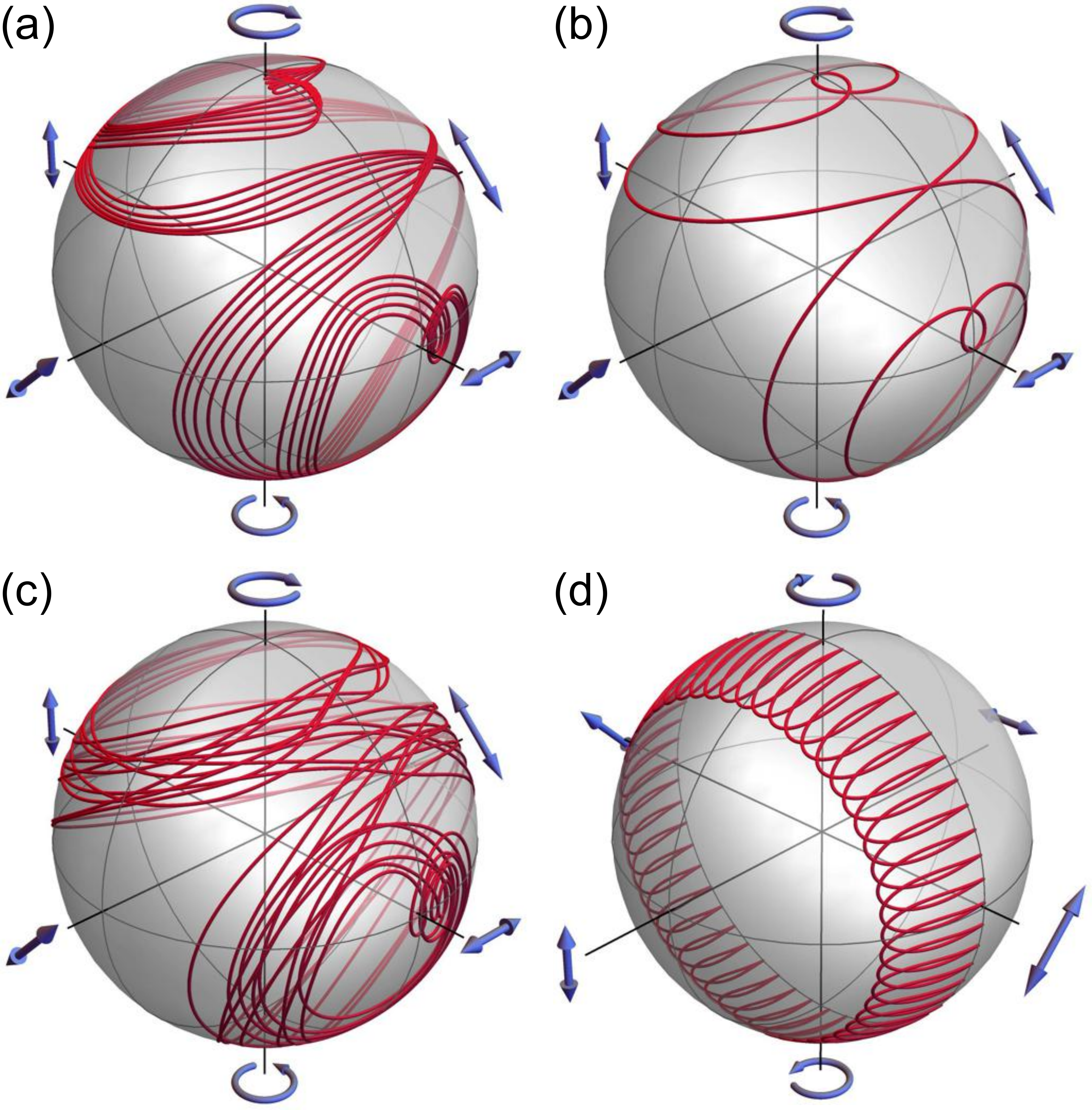}
\end{center}
\caption{ \label{SUPPL_figStokes}
The trajectories of the normalized Stokes vector $\vec{S}$ on the Poincar\'{e} sphere for the conditions treated in Fig.~3 in the main text (a), Fig.~\ref{SUPPL_figSphereRES01} (b), Fig.~\ref{SUPPL_figSphere02} (c) and Fig.~\ref{SUPPL_figSphere03} (d).
The polarization is taken at the angular coordinate~$\theta = 0$.
}
\end{figure}

The resonant case $\Omega = \omega_c \Delta _1$ is discussed in the main text where Fig.~3 illustrates the dynamics of the states occupancies and behaviour of the polariton polarization.
Figure~\ref{SUPPL_figSphereRES01} illustrates the dynamics of the states in the case when the nonresonant terms $A_{R,\downarrow}$ are neglected.
Figures~\ref{SUPPL_figStokes}(a) and~\ref{SUPPL_figStokes}(b) show trajectories of the Stokes vector on the Poincar\'{e} sphere calculated at the angular coordinate~$\theta=0$ in the resonant case  with (a) and without (b) nonresonant terms.
Comparing the figures, one can see that the nonresonant terms are responsible for breaking periodicity of the evolution of the polariton modes and making it quasi-periodical.

\begin{figure*}[tb]
\begin{center}
 \includegraphics[width=.8\linewidth]{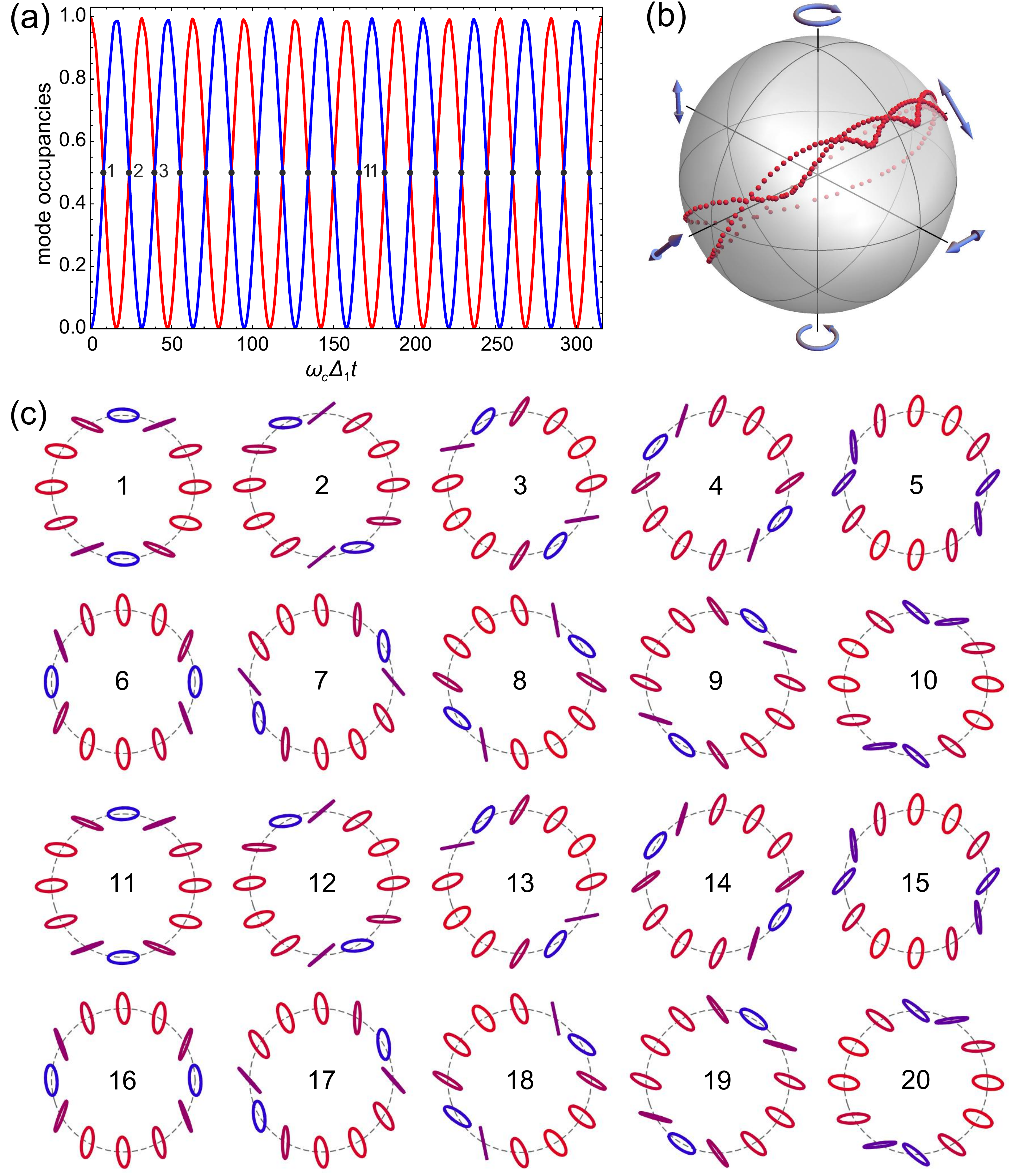}
\end{center}
\caption{ \label{SUPPL_figRotation}
The dependence of the occupancies of the resonant $A_{T,\uparrow}$ modes on time for the case of the exact resonance, ${\omega_c \Delta_1 = \Omega}$~(a).
Red dots on (b) show the polarization vectors $\vec{S}$  on the Poincar\'{e} sphere calculated at $\theta = 0 $ for the times corresponding to $|A_T|^2 = |A_{\uparrow}|^2$.
The polarization ellipses at different point of the ring for the indicated times are shown in~(c) numbered from 1 to 20 in accordance with the order of the intersections of the curves in~(a).
}
\end{figure*}

Let us note that after a beating period the field reproduces  itself (or quasi-reproduces in case when the none-resonant terms are accounted) in the rotating reference frame. This means that in the laboratory frame the filed distributions will be turned by the angle $\varphi=\Omega T_b$. Thus, if we plot the polarization ellipses at the times when the occupancies $\rho_T$ and $\rho_{\uparrow}$ becomes equal, see Fig.~\ref{SUPPL_figRotation},  the shift becomes obvious. The first and the elevens pictures looks very much similar because $\frac{\Omega T_b}{2\pi}$ is close to an integer. However in a general case the ratio $\frac{\Omega T_b}{2\pi}$ is irrational and thus the polarization state is never the same in the laboratory reference frame even if the non-resonant terms are neglected and the motion is periodic in the rotating reference frame. 

\begin{figure}[tb]
\begin{center}
 \includegraphics[width=\linewidth]{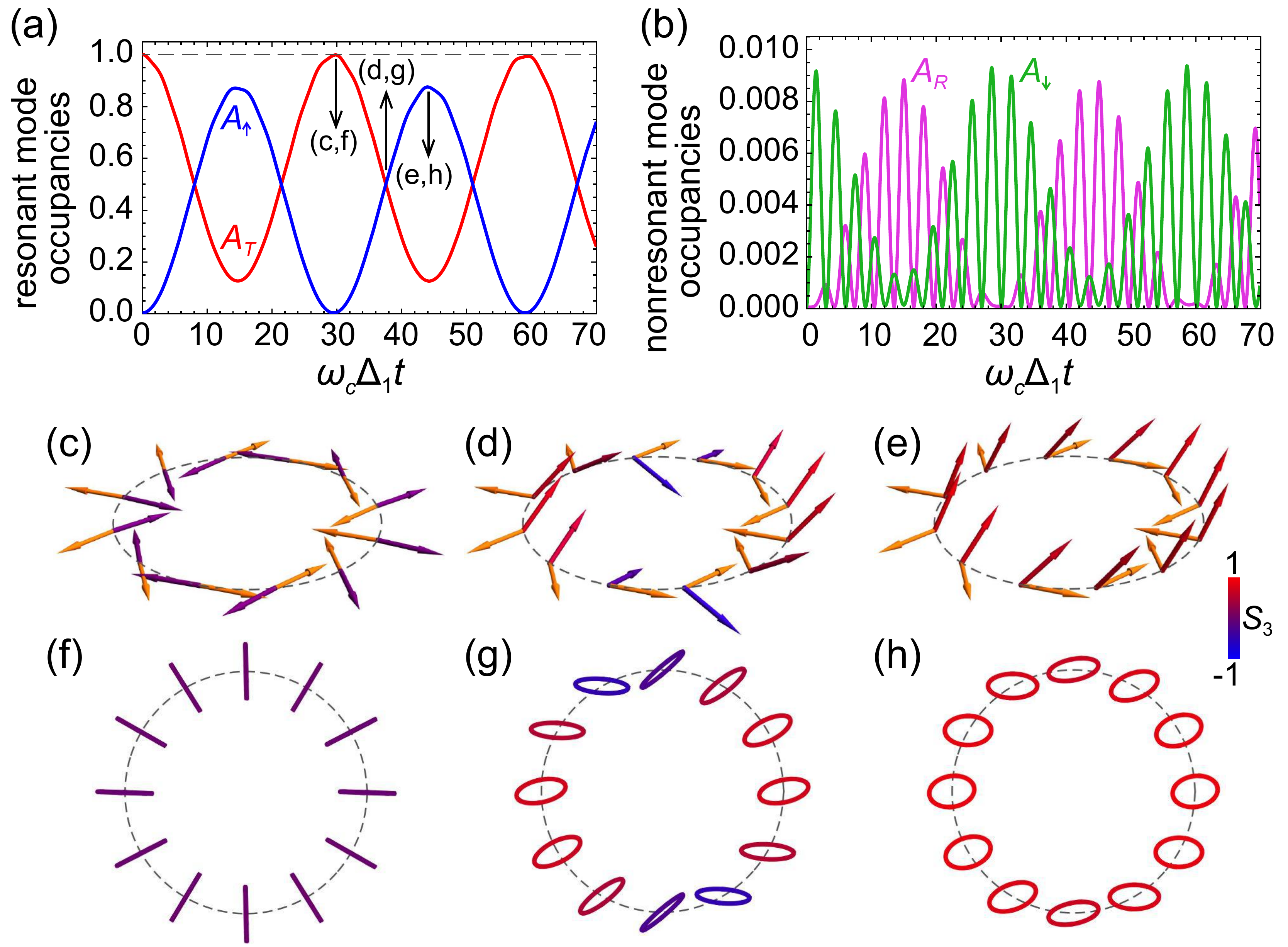}
\end{center}
\caption{ \label{SUPPL_figSphere02}
(Color online) The dependencies of the occupancies of the resonant $A_{T,\uparrow}$ (a) and nonresonant $A_{R,\downarrow}$ (b) modes on time for the small detuning from the resonance, $\Omega=1.075$.
The normalized Stokes vector $\vec{S}$ and the effective magnetic field at different angular positions on the ring are shown in panels (c)--(e) for the times indicated in (a).  The arrows  showing the Stokes vector are shown in color ranging from red to blue. 
Orange arrows indicate the orientation of the effective magnetic field. The polarization ellipses are shown in panels (f)--(h).
The parameters are $\omega_c \Delta_1=1$, $\eta=0.1$.
The initial conditions vector is 
${[A_{\text{T}}(0),A_{\text{R}}(0),A_{\uparrow}(0),A_{\downarrow}(0)] = (1,0,0,0)}$.
}
\end{figure}

In the nonresonant case $\Omega \ne \omega_c \Delta _1$, as it is seen from (\ref{eigenstates_simpl_occup2}) the occupancy of the circularly polarized state $\rho_{\uparrow}$ never reaches unity even if the non-resonant terms are disregarded. Correspondingly, the occupation of the tangential polarized state is always finite, see (\ref{eigenstates_simpl_occup1}). The dynamics of the occupancies is shown in Fig.~\ref{SUPPL_figSphere02}(a) for the finite detuning from the resonance.
Let us remark that the occupancy evolution observed in the numerical simulations fits perfectly to the one predicted by~\eqref{eigenstates_simpl12}.

It is worth mentioning here that in this case the Stokes vector trajectory never riches the pole (where the polarization is circular) but instead orbiting around the pole, see Fig.~\ref{SUPPL_figStokes}(c). The diameter of the orbit growth with the detuning from the resonance. The Stokes vector and the effective magnetic field at different points of the trap are shown in panels~\ref{SUPPL_figSphere02}(c--e) for the times when most of the polaritons in the linearly~(c) or circularly~(e) polarized mode and when the occupancies of the states are equal~(d). The polarization ellipses are shown in panels~\ref{SUPPL_figSphere02}(f--h) correspondingly.

\section{Energy splitting at low rotation velocities}

Here we consider the case when the rotation velocity is small $\Omega \ll \omega_c \Delta_1$. Then in the absence of the potential $\eta=0$ there are two energy levels $\varepsilon = \hbar \omega_c (1 \pm \Delta_1)$ corresponding to the radially and tangentially polarized modes $A_{R, T}$ and double degenerate energy level $\varepsilon=\hbar \omega_c$ corresponding to the circularly polarized modes $A_{\uparrow, \downarrow}$. At finite $\eta$ the degeneracy is lifted due to hybridization of the modes $A_{\uparrow, \downarrow}$.   

Let us consider this hybridization in detail. We focus of the modes having low egenergies  $\frac{\varepsilon}{\hbar} - \omega_c \ll {\omega_c \Delta_1}$. To insure small energy splitting we require that $\eta \ll \omega_c \Delta_1$. Then the amplitudes $b_1$ and $b_2$ in (\ref{eq:Floquet_spectrum}) are non-resonant and can be expressed as 
\begin{subequations}
\label{SEQb1b2lowrot}
\begin{eqnarray}	
&&b_{1}= - \frac{ \eta }{\omega_c \Delta_1}  \left( b_{3} + b_{4}\right), \, \\
&&b_{2}= \frac{\eta}{\omega_c \Delta_1} \left(  b_{3} - b_{4}\right). \, 
\end{eqnarray}
\end{subequations}
Substituting~\eqref{SEQb1b2lowrot} into (\ref{eq:Floquet_spectrum}c,d) we obtain
\begin{subequations}
\begin{eqnarray}
\label{eq:Floquet_approx_Omega0_1}
&&\frac{\varepsilon}{\hbar} b_{3}=(\omega_c -\Omega) b_{3} + \frac{2\eta^2 }{\omega_c \Delta_1} b_{4}, \,  \\
\label{eq:Floquet_approx_Omega0_2}
&&\frac{\varepsilon}{\hbar} b_{4}=(\omega_c +\Omega) b_{4} + \frac{ 2\eta^2 }{\omega_c \Delta_1} b_{3}. \,
\end{eqnarray}
\end{subequations}
The dependencies of the eigenenergies of the hybridized modes are given by 
\begin{eqnarray}
\varepsilon=\hbar \omega_c \pm \hbar \sqrt{ \Omega^2 + \frac{ 4\eta^4 }{\omega_c^2 \Delta_1^2} }.
\end{eqnarray}

The dynamics of the occupancies of the first $\rho_{\uparrow}$ and the second $\rho_{\downarrow}$ circularly polarized states can easily be calculated and the expressions for these quantities are
\begin{subequations}
\begin{eqnarray}
\label{eq:Om0_occup1}
&&\rho_{\uparrow}=1-\frac{4\eta^4}{\omega_c^2 \Delta_1^2 \Omega^2 + 4\eta^4 } \sin^2 \left (\sqrt{ \Omega^2 + \frac{ 4\eta^4 }{\omega_c^2 \Delta_1^2}} t \right), \quad \qquad  \\
\label{eq:Om0_occup2}
&&\rho_{\downarrow}=\frac{4\eta^4}{\omega_c^2 \Delta_1^2 \Omega^2 + 4\eta^4 } \sin^2 \left (\sqrt{ \Omega^2 + \frac{ 4\eta^4 }{\omega_c^2 \Delta_1^2}} t \right) \,
\end{eqnarray}
\end{subequations}
for the case when at $t=0$ only one circularly polarized state is populated with the occupancy equal to unit, ${\rho_{\uparrow}=1}$, ${\rho_{\downarrow}=0}$.

From~(\ref{eq:Om0_occup1}) it is seen that the beating period between the circularly polarized states becomes shorted for the higher coupling strength $\eta$ and the rotation frequency~$\Omega$: $T_b={\pi} \left( \Omega^2 + { 4\eta^4 }/{\omega_c^2 \Delta_1^2}\right)^{-1/2}$. The modulation depth, however, decreases with $\Omega$ and thus for the relatively rapidly rotating potential the polariton transfer from one circularly polarized state to another is suppressed. 

\begin{figure}[tb]
\begin{center}
 \includegraphics[width=\linewidth]{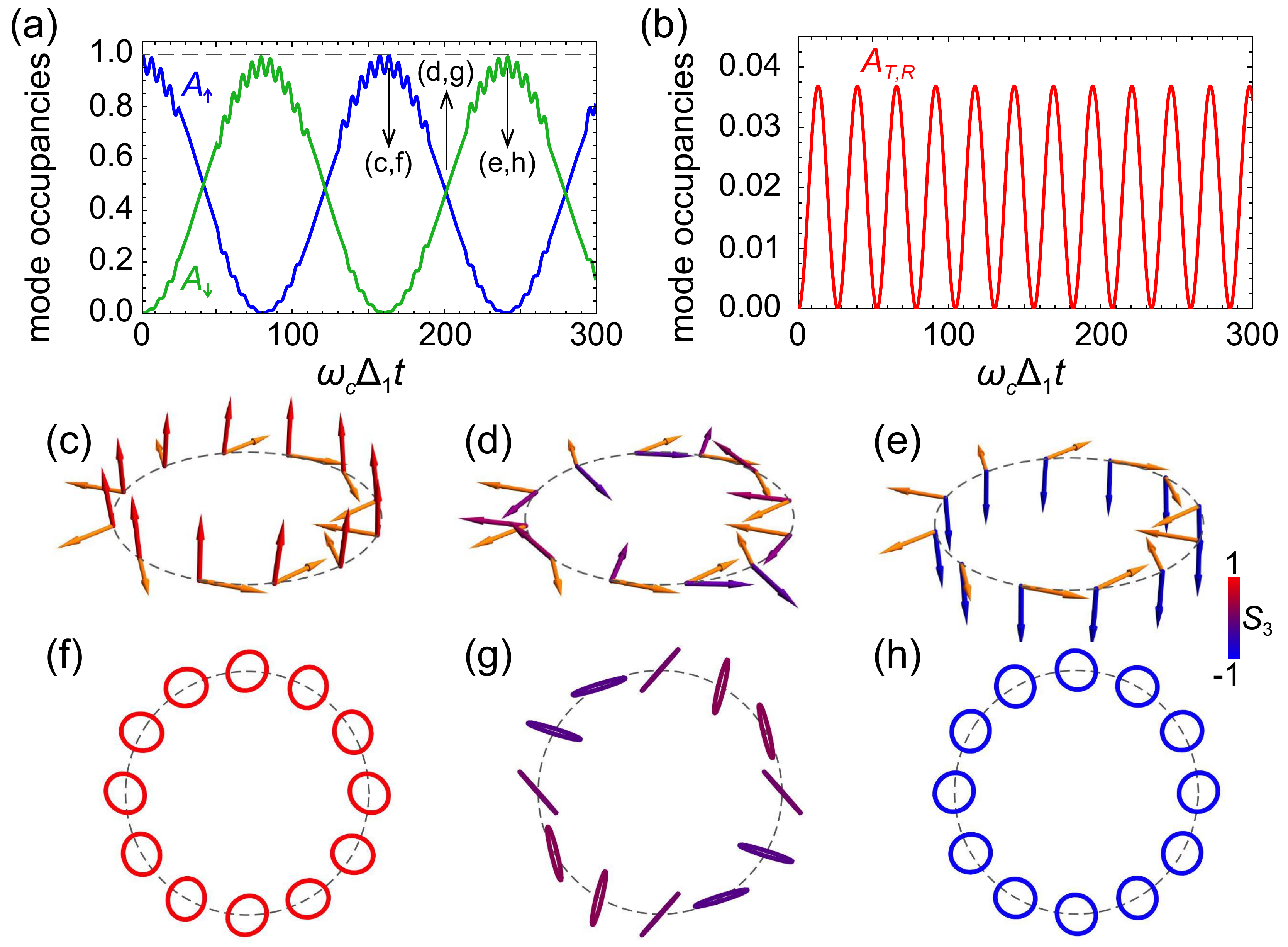}
\end{center}
\caption{ \label{SUPPL_figSphere03}
(Color online) The same as Fig.~\ref{SUPPL_figSphere02}, but for resting potential ($\Omega=0$).
The initial conditions vector is 
${[A_{\text{T}}(0),A_{\text{R}}(0),A_{\uparrow}(0),A_{\downarrow}(0)] = (0,0,1,0)}$.
}
\end{figure}

The evolution of the occupancies $\rho_{\uparrow}$ and $\rho_{\downarrow}$ are shown in Fig.~\ref{SUPPL_figSphere03}(a) for the case $\Omega=0$ calculated for the case when the non-resonant terms are accounted. The polarization dynamics are illustrated in panels (c)--(h). The non-resonant terms are responsible for quasi-periodic dynamics. They also explain why the occupancies of the circularly polarized states $\rho_{\uparrow}$ and $\rho_{\downarrow}$ are never equal to~$1$ for~$t>0$. Decreasing the coupling strength $\eta$ the effect of the non-resonant terms can be reduced but this make the beating period large. One can anticipate that for the finite losses this can prevent the observation of the polarization beating.

The important fact which should be mentioned here is that in this case the mode interaction is mediated by the non-resonant terms and, therefore, the splitting at $\Omega=0$ is  $\frac{\hbar \eta^2 }{\omega_c \Delta_1}$ whereas at $\omega=\omega_c \Delta_1$ the splitting is $\hbar\eta$. So the splitting is proportional to square of $\eta$ in the former case and to $\eta$ in the latter case. Thus for the small $\eta$ the splitting is stronger for the potential rotating at the velocity $\Omega=\omega_c \Delta_1$.